\documentclass[11pt,onecolumn,legalpaper]{article} 

\usepackage{amsmath} 
\usepackage{amssymb,epsf}
\usepackage{xcolor}
\usepackage{soul}
\usepackage{float}
\usepackage[caption=false]{subfig}
\usepackage{wrapfig}
\usepackage{enumitem}
\usepackage{blindtext}
\usepackage{adjustbox}
\usepackage{tabularx}
\usepackage{graphicx}
\usepackage{txfonts}
\usepackage{lipsum}
\usepackage{authblk}
\usepackage{changepage}
\usepackage{tikz}
\usepackage{multicol}
\usepackage{geometry}
\usetikzlibrary{matrix}

\usepackage{natbib,twoopt}
\usepackage[breaklinks=true]{hyperref} 
\bibpunct{(}{)}{;}{a}{}{,} 

\title{Relativistic Atomic Structure of Au IV and the Os Isoelectronic Sequence: opacity data for kilonova ejecta}

\author{Z. S. Taghadomi$^{1}$\footnote{email address: zt33957@uga.edu},
	Y. Wan$^{1}$ \footnote{email address:
		yier.wan123@gmail.com} , A. Flowers$^{1}$\footnote{email
		address: acf55815@uga.edu} , P. C. Stancil$^{1}$\footnote{email
		address: pstancil@uga.edu}, B. M. McLaughlin$^{2}$\footnote{email
		address: b.mclaughlin@qub.ac.uk}, S. Bromley$^{3}$\footnote{email address: sjb0068@auburn.edu}, J. P. Marler$^{4}$\footnote{email address: jmarler@g.clemson.edu}, C. E. Sosolik$^{4}$\footnote{email address: sosolik@clemson.edu}, S. Loch$^{3}$\footnote{email
		address: lochstu@auburn.edu}} \affil{$^1$Department of Physics and Astronomy, Center for Simulational Physics, The University of Georgia, Athens, GA 30602, USA\\
		$^2$Centre for Theoretical Atomic and Molecular Physics (CTAMOP), School of Mathematics and Physics, Queen’s University Belfast, Belfast BT7 1NN, UK \\
		$^{3}$ 	Department of Physics and Astronomy, Auburn University, Auburn, AL 36849, USA\\
		$^{4}$ 	Department of Physics and Astronomy, Clemson University, 104 Kinard Laboratory, Clemson, SC 29634, USA }

\date{}

\begin{document}
 \maketitle

 \begin{abstract}
 
Direct detection of gravitational waves (GW) on Aug. 17, 2017, propagating from a binary
neutron star merger, opened the era of multimessenger astronomy. The ejected material from neutron star
mergers, or “kilonova”, is a good candidate for optical and near
infrared follow-up observations after the detection of GWs. The kilonova from the ejecta of GW1780817 provided the first evidence for the astrophysical site of the synthesis of heavy nuclei through the rapid neutron capture process or r-process. Since properties of the emission are largely affected by opacities of the ejected material, enhancements in the available, but limited r-process atomic data have been motivated recently. However, given the complexity of the electronic structure of these heavy elements, considerable efforts are still needed to converge to a reliable set of atomic structure data.
The aim of this work is
to alleviate this situation for low charge state elements in the Os-like isoelectronic sequence. In this regard, the general-purpose relativistic atomic structure packages (GRASP$^0$ and GRASP2K) were used to obtain
energy levels and transition probabilities (E1 and M1). We provide line lists and expansion opacities for a range of r-process
elements.
We focus here on the Os isoelectronic sequence (Os I, Ir II, Pt III, Au IV, Hg V). The results are benchmarked against existing experimental data and prior calculations, and predictions of emission spectra relevant to kilonovae are provided. Fine-structure (M1) lines in the infrared potentially observable by the James Webb Space Telescope are highlighted.


\end{abstract}

\begin{multicols*}{2}
\section{Introduction}
The usefulness of gravitational wave observations are now well established, where focus has been on the detection of merging black holes \citep{abbott2016observation}. However, one of the most promising applications of gravitational wave studies was revealed with the first detection of merging neutron stars and their role in producing heavy elements \citep{wanajo2014production}. It has been established that low-mass
elements C, N, O, ..., Mg, and up to Mo are produced in low-mass stars, while elements in the iron-peak are produced in the explosive ejecta of supernovae. Numerous observations and robust
nucleosynthesis models confirm this scenario \citep{nomoto2006nucleosynthesis,grimmett2018nucleosynthesis}. For heavier elements that require a rapid neutron capture (r-process) mechanism, the details of their production mechanism are an area of active study.\\

It was postulated as early as 1957 by Cameron that the merger of binary
neutron stars would result in a neutron rich ejecta that could produce these
heavy elements. One of the early nucleosynthesis models which predicted
the yields of such events was given by Freiburghaus et al. \citet{freiburghaus1999r}.
The detection of a neutron star merger (NSM) \citep{abbott2017gw170817} event and the spectral
observation of its optical and infrared emission opens up the possibility of
direct interrogation of the formation site of these heavy r-process elements.
However, interpretation, analysis, and modeling of the observed optical/IR spectra is hindered by the sparsity of atomic opacity studies for these r-process elements, particularly beyond Mo. \\

Our goal is to provide additional large-scale, but reliable atomic line lists for Re-Pt ($Z=75-79$). Thus atomic data, such as wavelengths, spectroscopic labels, and
transition probabilities are needed. While there are many computational methods based on perturbation or variational theory that could be adopted, we focus on general multiconfiguration variational methods where the wave function for an atomic state is determined in terms of a basis of configuration state functions (CSFs). In particular, the multiconfiguration Dirac-Hartree-Fock (MCDHF) method with Breit and quantum electrodynamics (QED) corrections \citep{kato1957eigenfunctions} as implemented in the General purpose Relativistic Atomic Structure Package (GRASP) \citep{parpia1996grasp92,jonsson2013new} is adopted. This fully relativistic code constructs and diagonalizes a Dirac-Coulomb Hamiltonian
to produce relativistic orbitals, which are then used in a multiconfiguration
Dirac-Fock calculation.\\

The atomic data that are needed for spectral and neutron star merger modeling includes wavelengths of emission lines (hence energies from the atomic structure), E1 and M1 spontaneous emission rates. It is also useful to produce derived coefficients that can be directly used by modelers. One particularly useful coefficient is known as an expansion opacity. In an expanding object, the frequency of the photons suffers a continuous Doppler shift with respect to the rest frame of the material. Thus each photon has an increased probability of interacting with a line; this enhanced effect of spectral lines can be taken into account as an expansion opacity \citep{karp1977opacity}.\\

In this manuscript we focus on the Os isoelectronic sequence (Os I ,..., Hg V), because it covers elements likely to be made in neutron star mergers and is confined to the lower charge states likely to be present. We note that work on other isoelectronic sequences is also underway. Section 2 describes the relativistic atomic structure theory adopted here, relations for emission spectra, and opacities in the local thermodynamic equilibrium (LTE) approximation. Section 3 gives a discussion of our results, while a summary and outlook for future work is given in Section 4.

\section{Atomic Structure Calculations: GRASP$^0$ and GRASP2K}
\subsection{GRASP}
The GRASP$^0$ package is based on the Oxford multiconfiguration Dirac-Fock (MCDF) \citep{grant1980atomic} and higher-order corrections (MCBP/BENA) \citep{mackenzie1980program} codes published in 1980. The package is used to compute atomic energy levels, orbitals, and transition data within the relativistic formalism.

The GRASP2K package is a later iteration of GRASP$^0$ and based on the fully relativistic multiconfiguration Dirac-Hartree-Fock (MCDHF) method \citep{grant2007relativistic}. The package consists of a number of application programs and tools to compute approximate
relativistic wave functions, energy levels, hyperfine structures (HFS), isotope shifts (IS), Lande
g-factors, interactions with external fields, angular couplings for labeling purposes, transition energies, and transition probabilities for many-electron atomic systems. We use both of these codes as a means of checking the convergence of the calculations and to give an indication of the approximate uncertainties in the final results.
\subsection{Multiconfiguration Dirac-Hartree-Fock (MCDHF)} 
 In the relativistic multiconfiguration Dirac-Hartree-Fock (MCDHF) method, the Dirac-Coulomb Hamiltonian is given by,
\begin{equation}
H_{DC}=\sum_{i=1}^{N} [c\alpha_{i}.p_{i}+(\beta_{i}-1)c^{2}+V_{nuc}(r_{i})]+\sum_{i>j}^{N} \frac{1}{r_{ij}},\label{eq:hdc}
\end{equation}
where $V_{nuc}$ is the electron-nucleus coulomb interaction, $r_{ij}$ is the distance between electrons $i$ and $j$, and $\alpha$ and $\beta$ are the Dirac matrices. The wave function is represented by an atomic state function (ASF), $\psi(\gamma P J)$, expanded in a set of configuration state
functions (CSFs), $\Phi(\gamma P J)$,
\begin{equation}
\psi(\gamma P J)=\sum_{j=1}^{N_{CSF}} c_{j}\Phi(\gamma_{j} P J).\label{eq:wf}
\end{equation}
Here $\gamma$ specifies CSF properties including orbital occupancy and subshell quantum numbers in the angular momentum coupling tree, $P$ is parity, $J$ is the final angular quantum number, and $c_{j}$ are the expansion coefficients. The CSFs can be expressed as products of one-electron solutions of the Dirac equation which can be written as,
\begin{equation}
\psi_{nlsjm}(r,\theta,\phi)=\frac{1}{r}\begin{pmatrix} 
P_{nlj}(r)\Omega_{lsjm}(\theta,\phi)  \\
iQ_{nlj}(r)\Omega_{\tilde{l}sjm}(\theta,\phi) 
\end{pmatrix},\label{eq:wf1e}
\end{equation}
where $P_{nlj}(r)$ and $Q_{nlj}(r)$ are called the regular and irregular solutions, and $\Omega_{lsjm}(\theta \phi)$ are built from the coupling of the spherical harmonics $Y_{lm_{l}}(\theta,\phi)$ and the spin functions $\chi_{m_{s}}^{(1/2)}$. $n$, $l$, $j$, and $s$, are the usual principal, orbital angular momentum, total angular momentum, and spin angular momentum quantum numbers, respectively, while $m_l$ is the projection of $l$ and $m_s$ the projection of $s$ on the $z$ quantization axis. 
$l$ and $\tilde{l}$ are related to each other \citep{greiner2000relativistic},

\begin{equation}
\tilde{l}=\left\{
\begin{array}{lr} 
\ l+1,   &  j=l+1/2 \\
\ l-1,  &  j=l-1/2.
\end{array}\right.\label{eq:l}
\end{equation}
If the quantum number $\kappa$ is introduced as the eigenvalue of the
operator $K=1-l\sigma$, where $\sigma$ is the Pauli matrix,
\begin{equation}
\kappa=\left\{
\begin{array}{lr}
\ -(l+1),   &  j=l+1/2\  (\kappa\ negative) \\
\ +l,  &  j=l-1/2\  (\kappa\ positive),
\end{array}\right.\label{eq:k}
\end{equation}
then we can rewrite Eq. {\eqref{eq:wf1e}} as,
\begin{equation}
\psi_{n \kappa m}(r,\theta,\phi)=\frac{1}{r}\begin{pmatrix} 
P_{n \kappa}(r)\Omega_{\kappa m}(\theta,\phi)  \\
iQ_{n \kappa}(r)\Omega_{-\kappa m}(\theta,\phi) 
\end{pmatrix}.\label{eq:wf2}
\end{equation}
The lists of CSFs that defines the ASF, mixing coefficients, and orbital parts of the wave functions are the basis for the GRASP2K program. Thus, the choice of CSFs affects the accuracy of the calculated energies and transition probabilities. In the work to be presented here we perform GRASP$^0$ and GRASP2K computations with a range of CSFs, as described in Section 3.

\subsection{LTE Spectra and Expansion Opacity} 
The environment in the kilonova ejecta at early times is expected to be of a sufficiently high density that the populations of the electronically excited states are considered to be thermalized, i.e., the ejecta gas density exceeds the critical density of the individual levels. The critical density is the gas density required to drive the populations of the individual levels to their thermodynamic values. Therefore, we can consider the emission or absorption to be in local thermodynamic equilibrium (LTE). In this regard, the line intensities are obtained from,
\begin{equation}
I(i,k)=F(i,k)A_{ki},\label{eq:I}
\end{equation}
where $A_{ki}$ is the  transition probability,  $f_{ik}$, the  absorption oscillator strength, and $F(i,k)$, the relative intensity, are obtained from,
\begin{equation}
f_{ik}=\frac{\lambda^2}{6.6702\times 10^{15}}\frac{g_{k}}{g_{i}}A_{ki},\label{eq:f}
\end{equation}
where $g_i$ and $g_k$ are the statistical weights for the lower and upper level of a spectral line which are obtained from the appropriate angular momentum quantum numbers, $g_{i(k)}=2J_{i(k)}+1$, and
\begin{equation}
F(i,k)=(2J_i+1)\exp(-\Delta E/kT)/Q,\label{eq:F}
\end{equation}
with $\Delta E=E_k-E_i$ and the internal partition function, Q,
\begin{equation}
Q=\sum_{i=1}(2J_i+1)\exp(-\Delta E/kT).\label{eq:Q}
\end{equation}

Following Kasen et al. \citep{kasen2013opacities}, the expansion opacity for bound-bound transitions is given by
\begin{equation}
\kappa_{\rm ex}(\lambda) = \frac{1}{ct_{\rm ej} \rho} \sum_{i, k} \frac{\lambda_{ik}}{\Delta \lambda}
[1-\exp( -\tau_{ik})], \label{eq:opacity}
\end{equation}
where the optical depth $\tau_{ik}$ is 
\begin{equation}
\tau_{ik} = \frac{\pi e^2}{m_e c} f_{ik} n_i t_{\rm ej} \lambda_{ik}.
\end{equation}
$n_i$ is the number density in the lower state of the ion in question, $t_{\rm ej}$ is the time since the merger, and $\rho$ is the ejecta density. From Kasen et al. \citep{kasen2013opacities}, we adopt the typical values $\rho=10^{-13}$ g cm$^{-3}$, $t_{\rm ej}$ = 1 day, and the wavelength binning $\Delta \lambda = 0.01 \lambda$. Note that these expansion opacities are provided for illustrative purposes showing the use of the fundamental data and to allow us to explore the effects of the atomic data on opacities. We assume the gas to be completely composed of only one ion. The data user can also generate their own opacities from the fundamental data including relevant mixtures of ions.

\section{Results and Discussions}
\subsection{Energy Levels and Transition Probabilities}
We have obtained a set of results for the energy levels and transition probabilities for the Os isoelectronic sequence using the GRASP$^0$ and GRASP2K codes. NIST Atomic Spectral Database \citep{NIST_ASD} has energies for some of the Os I \citep{moore1958atomic} and Ir II \citep{van1978term} levels. Also, some calculations have been done by \citet{gillanders2021constraints} for Pt III, and the HULLAC code has been adopted to compute atomic data for for Os I, Ir II, Pt III, and Au IV \citep{tanaka2020systematic}, but only for E1 transitions. \\

Our choice of adopted electron configurations is given in Table \ref{target} which includes the ground configuration and the configurations likely to dominate the expansion opacity under the low temperature radiation field conditions of the kilonova. This means that we focused on low-lying configurations. Thus, in the GRASP2K calculation we included $5d^6 6s^2, 5d^6 6s6p, 5d^7 6s, 5d^7 6p,$ and $5d^8$. In the GRASP$^0$ calculations, we performed three different computations with increasingly larger numbers of configurations to allow for convergence checks, with the configurations for the three calculations (10, 12, and 16 configurations) being shown in Table \ref{target}. However, it was found that convergence could be obtained in the GRASP2K calculations with smaller configuration sizes.\\

\begin{figure}[H]
	\begin{adjustwidth}{-1.5cm}{}
	\centering
	\subfloat[]{\includegraphics[width=10.5cm]{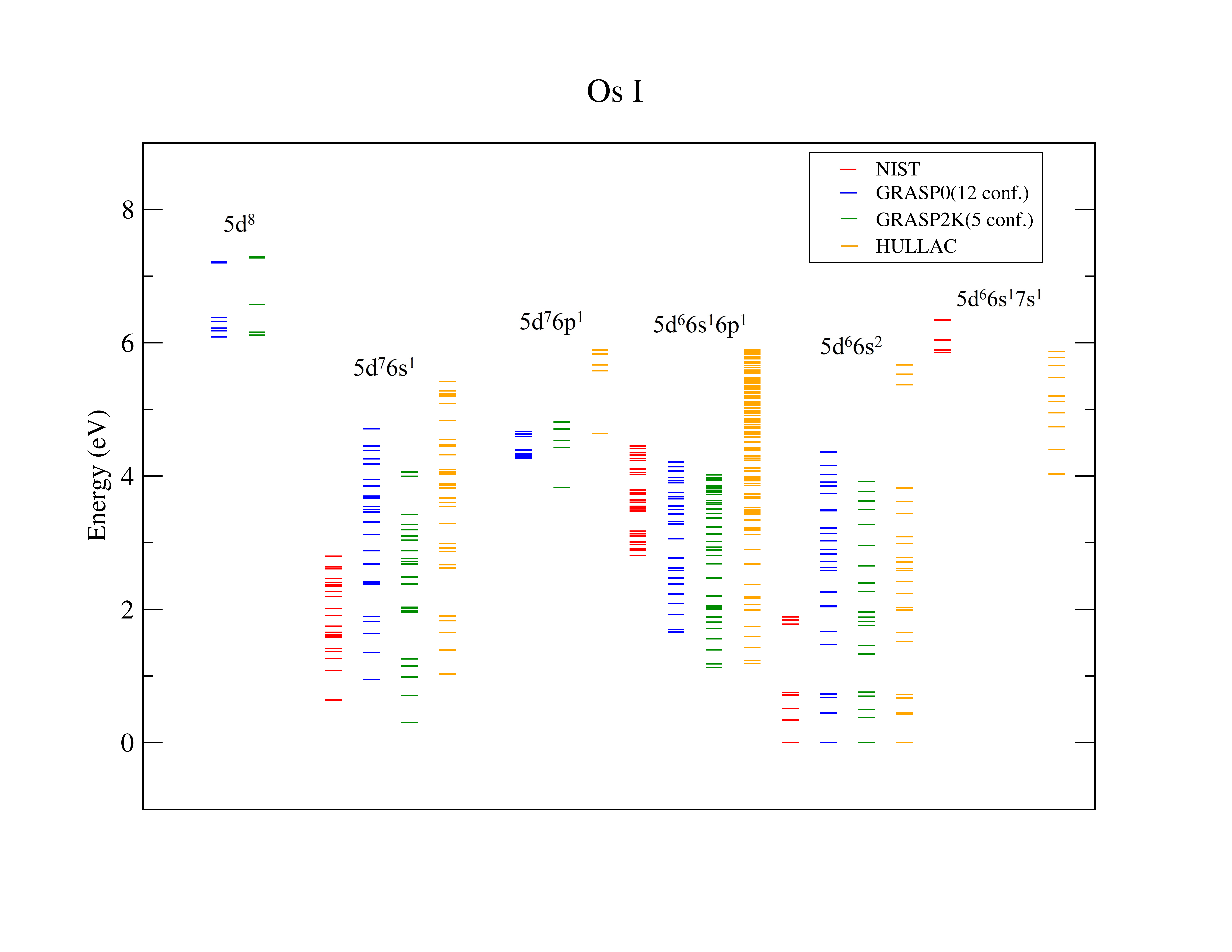}
		\label{OsIenergy}}	
	\subfloat[]{\includegraphics[width=10.5cm]{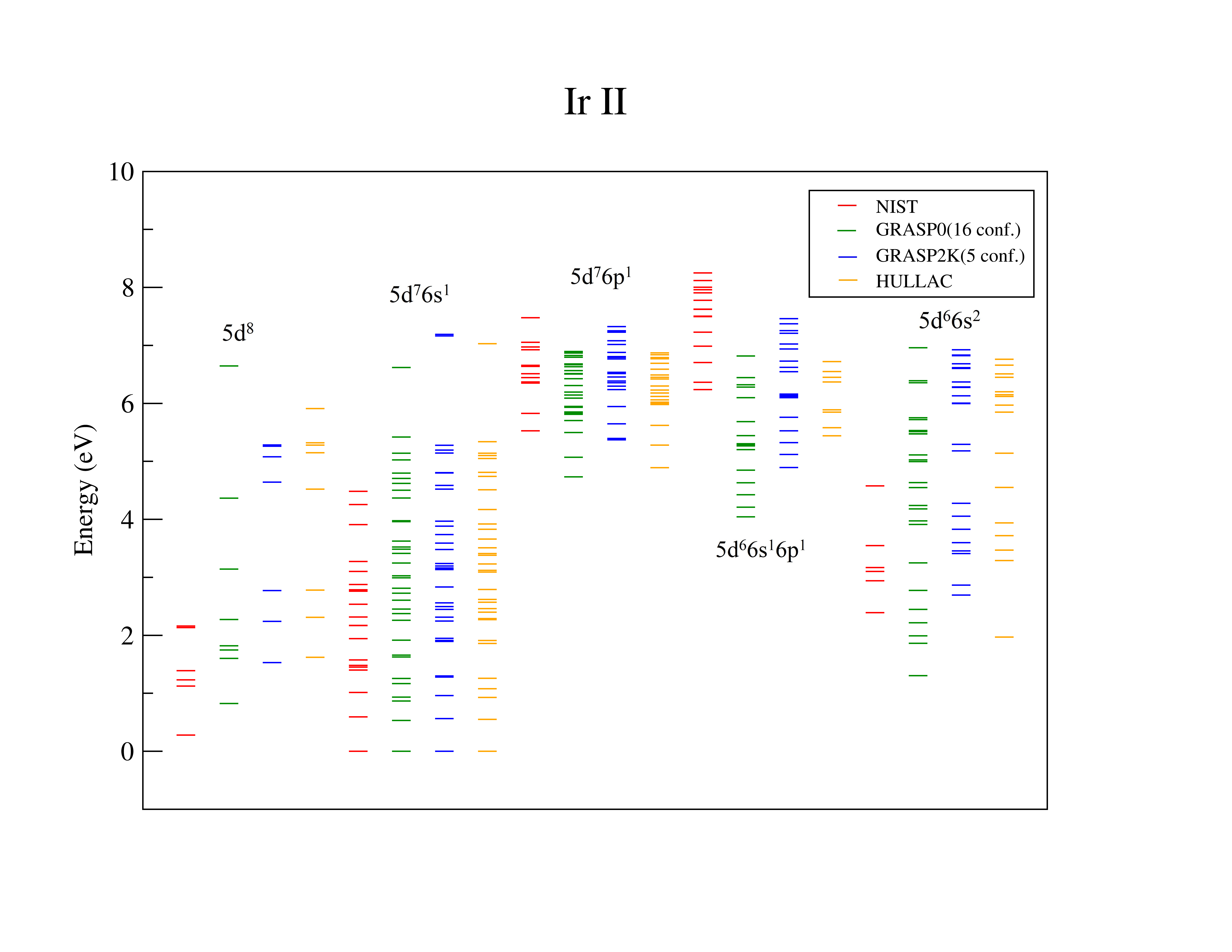}
		\label{IrIIenergy}}	
	\hfill
	\subfloat[]{\includegraphics[width=10.5cm]{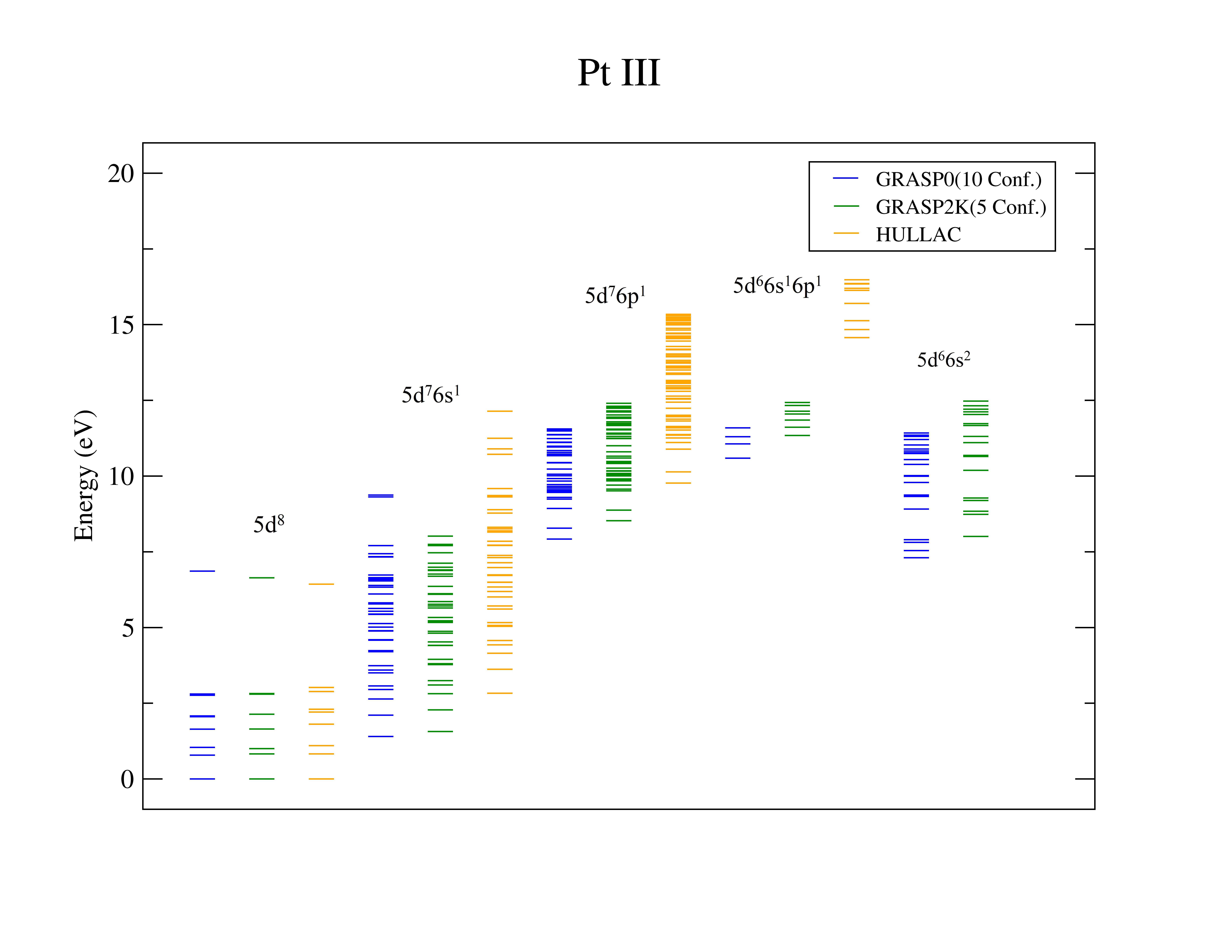}
		\label{PtIIIenergy}}
	\subfloat[]{\includegraphics[width=10.5cm]{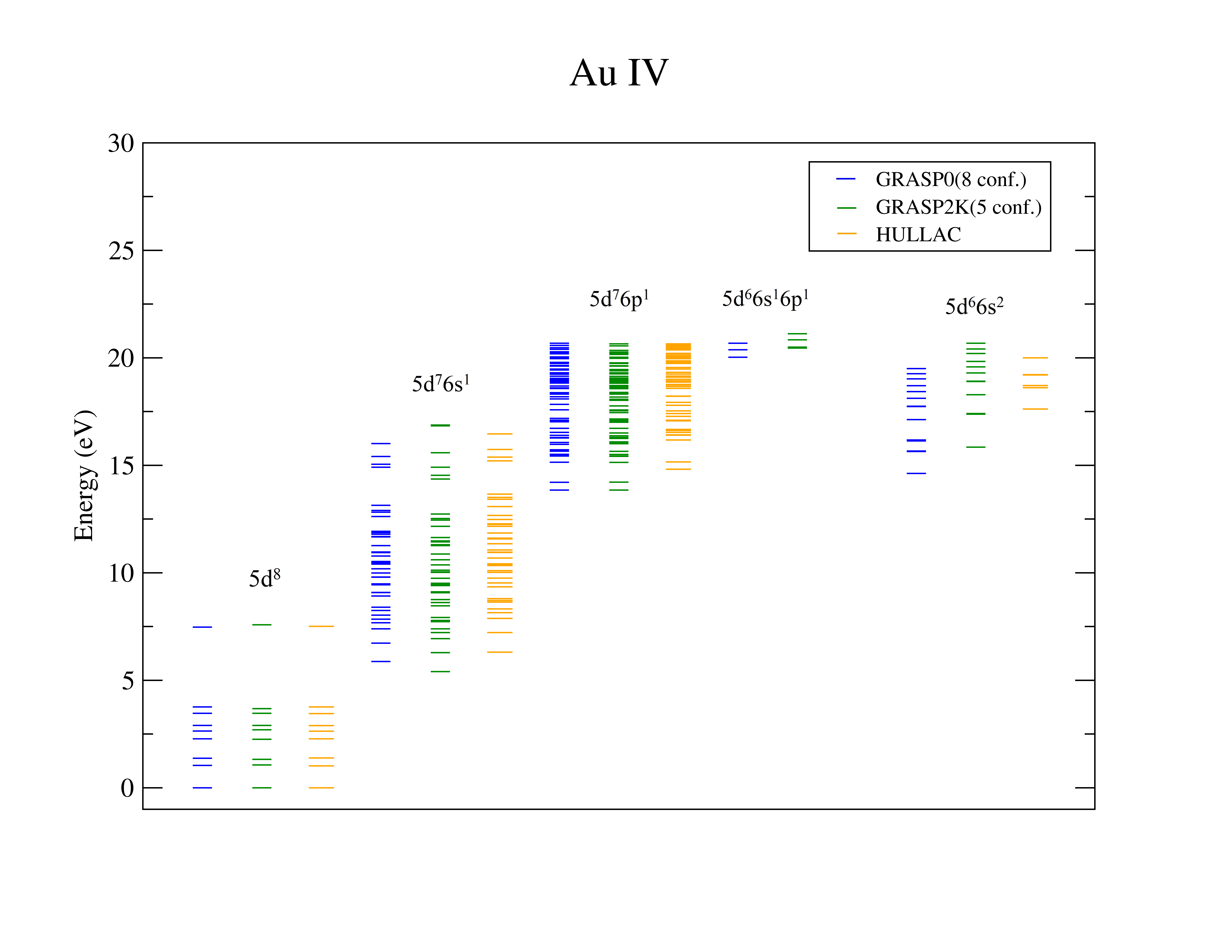}\label{AuIVenergy}}	
		\hfill	
		\hspace*{3cm}
	\subfloat[]{\includegraphics[width=10.5cm]{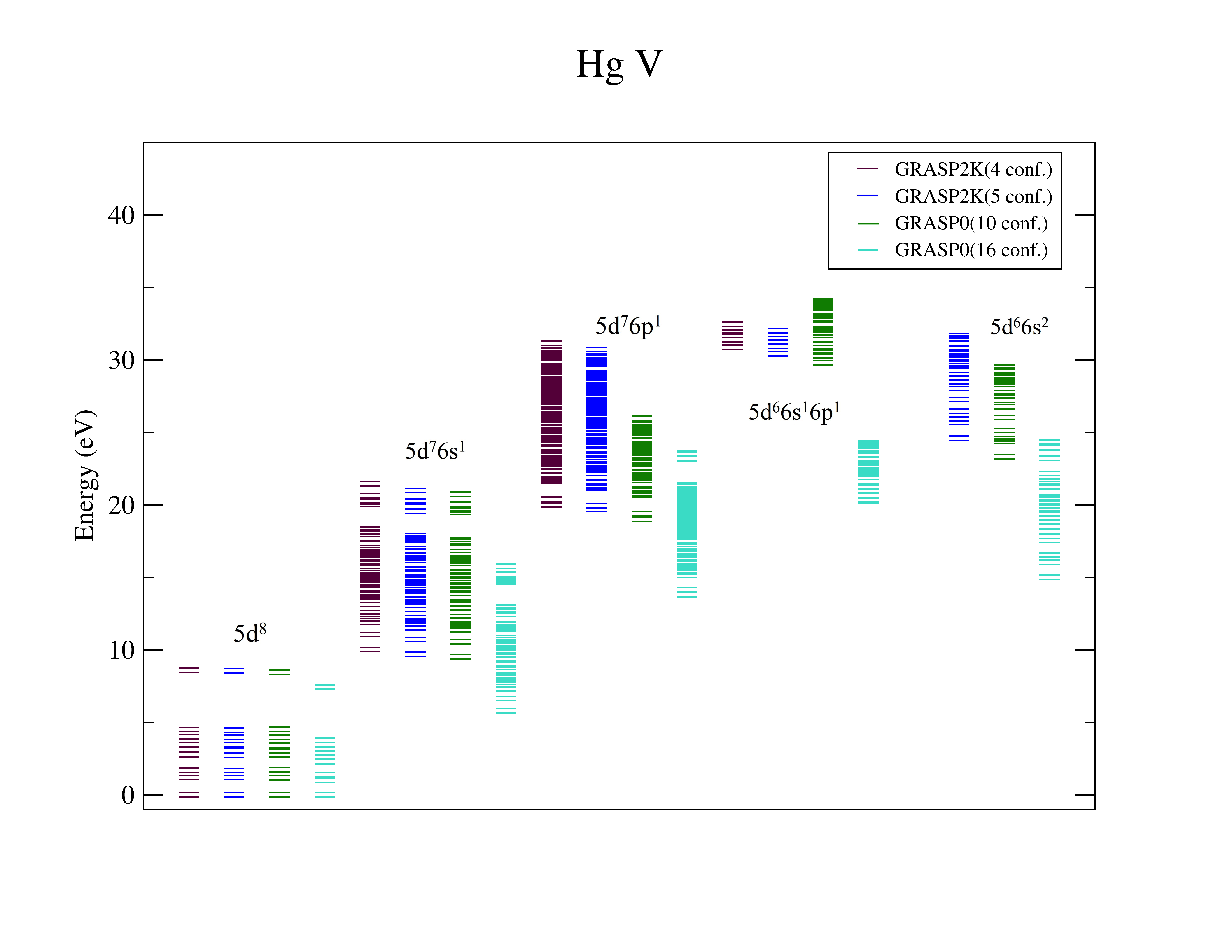}
		\label{HgVenergy}}
		\end{adjustwidth}
	\caption{Energy level diagrams for a) Os I, b) Ir II, c) Pt III, d) Au IV, and e) Hg V, comparing our results to available NIST data \citep{NIST_ASD} and HULLAC calculations \citep{tanaka2020systematic}. Each stack of levels refers to excited states from the indicated configuration.}\label{energy}
         \end{figure}

Figure \ref{energy} shows energy level diagrams for \ref{OsIenergy}) Os I, \ref{IrIIenergy}) Ir II, \ref{PtIIIenergy}) Pt III, \ref{AuIVenergy}) Au IV, and \ref{HgVenergy}) Hg V obtained from GRASP2K and GRASP$^0$ calculations using the target model in Table \ref{target} as the reference configurations. Comparison is made to theoretical data calculated using the HULLAC code \citep{tanaka2020systematic} and experimental data from the NIST database when available \citep{NIST_ASD}. Moreover, in GRASP2K
we used 4 and 5 reference configurations; for 4 configurations, the $5d^66s^2$ was omitted. Since the ground level is $5d^66s^2$ for Os I, the $5d^8$ was omitted in that case. In general, the GRASP2K energies are in better agreement with the NIST values than the GRASP$^0$ energies for the lower lying configurations.  The GRASP$^0$ calculations show some improvement in the energies for certain levels, while in most cases the comparison is mixed. As a consequence, we adopt the GRASP2K results for the expansion opacities later in this paper.  \\

The energy levels of Os I and Ir II are given in Table \ref{tabosi} and Table \ref{tabirii}, respectively, and are compared with energies from GRASP$^0$, NIST \citep{NIST_ASD}, HULLAC calculations \citep{tanaka2020systematic}, DESIRE \citep{fivet2007transition}, and experimental studies of Ir II \citep{xu2007improved}.  Although levels in the ground term are in good agreement with the NIST and DESIRE data, there are deviations from the NIST database for both GRASP2K and GRASP$^0$ results for the higher energy levels. For Pt III, Au IV, and Hg V, literature on the energy level structures is sparse. For Pt III, a GRASP$^0$ calculation is available  \citet{gillanders2021constraints}. For the remainder of the Os-sequence, we are limited to comparisons between the present GRASP$^0$ and GRASP2K calculations.  \\

Table \ref{energycomparison} gives as an example the convergence of energy eigenvalues for levels in the ground term of each of the considered ions from the current calculations. Here we give energies for all GRASP$^0$ and GRASP2K results. In all cases, the ground level is correctly predicted and the ordering of $J$-levels is consistent, except for Pt III from GRASP$^0$. \\

Computed energies for low-lying levels of Pt III are compared in Table \ref{tabptiii} from our GRASP2K(5) calculations to those of Gillanders et al. \citep{gillanders2021constraints} and the HULLAC calculations of Tanaka et al. \citet{tanaka2020systematic}, while comparison to the two M1 transition probabilities given in \citet{gillanders2021constraints} are also shown. There is general agreement amongst the calculations except for the energy of the $^5D_4$ state, and our M1 values are $36-52\%$ smaller than those computed by Gillanders et al. \citep{gillanders2021constraints}. \\

\begin{figure}[H]
	\begin{adjustwidth}{-1.5cm}{}
	\centering
	\subfloat[Os I]{\includegraphics[width=10.5cm]{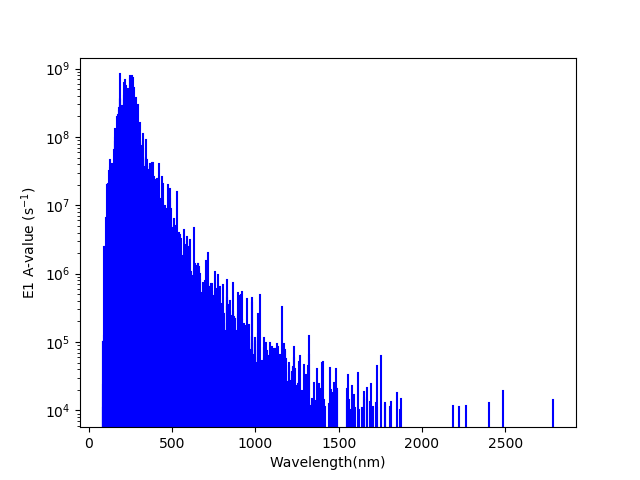}
		\label{OsIE1}}	
	\subfloat[Ir II]{\includegraphics[width=10.5cm]{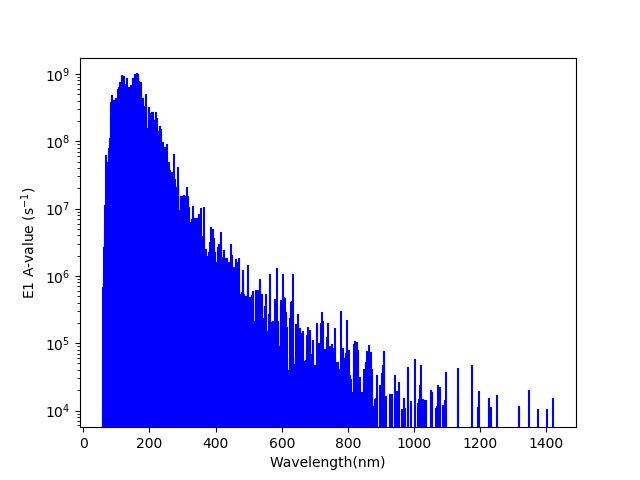}
		\label{IrIIE1}}
	\hfill
	\subfloat[Pt III]{\includegraphics[width=10.5cm]{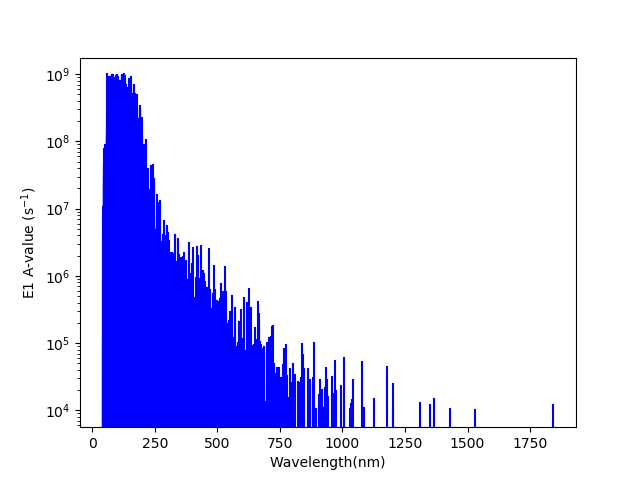}\label{PtIIIE1}}
	\subfloat[Au IV]{\includegraphics[width=10.5cm]{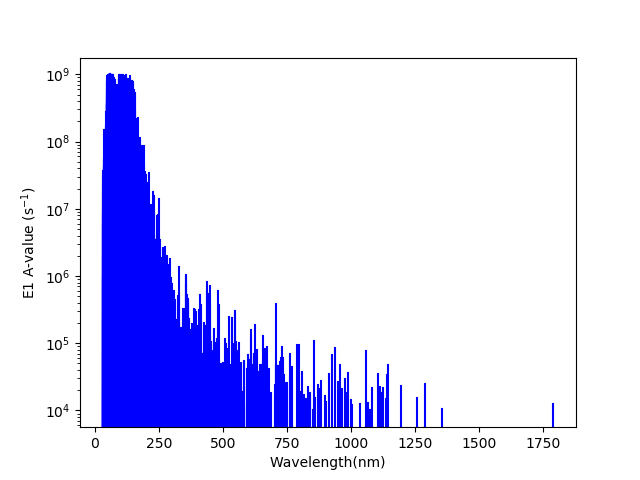}\label{AuIVE1}}
	\hfill	
	\hspace*{3cm}
	\subfloat[Hg V]{\includegraphics[width=10.5cm]{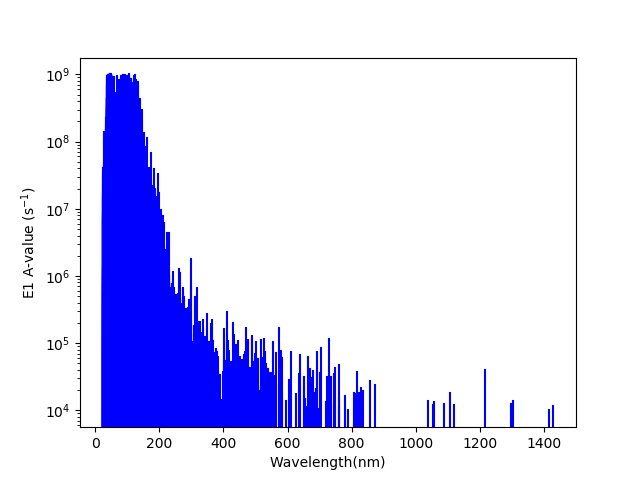}
		\label{HgVE1}}
		\end{adjustwidth}
	\caption{Transition probabilities (E1) for a) Os I, b) Ir II, c) Pt III, d) Au IV, and e) Hg V.}\label{E1}
\end{figure}

\begin{figure*}
	\centering
	\includegraphics[width=12cm]{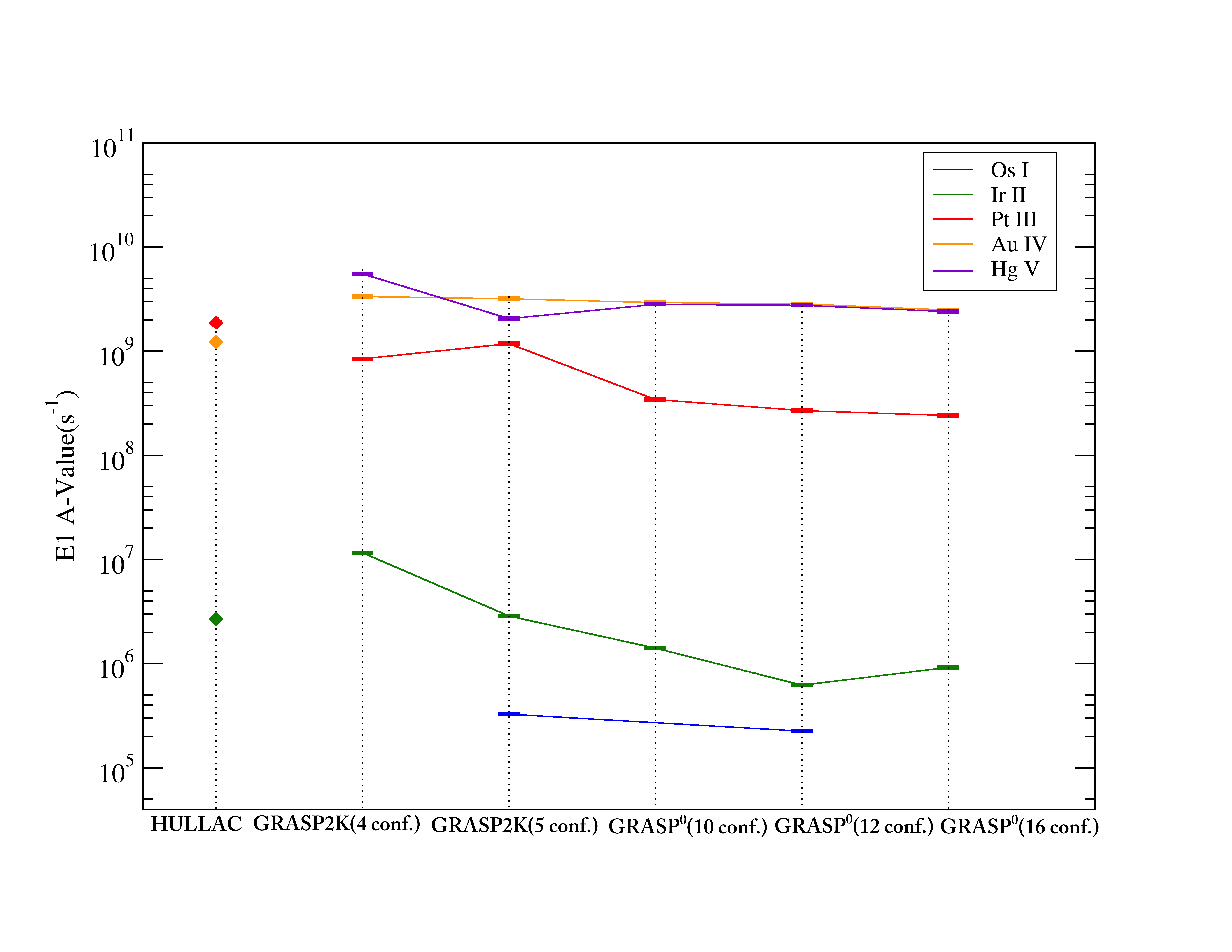}
	\caption{Convergence of E1 transition probability for different methods and numbers of configurations for the $5d^8$  $^3F$ $\rightarrow$ $5d^7(^2H_3)6p$  $^3G$ transition.}
	\label{convergence}
\end{figure*}

Transition probabilities for E1 vs. wavelength are plotted in Figure \ref{E1} for \ref{OsIE1}) Os I, \ref{IrIIE1}) Ir II, \ref{PtIIIE1}) Pt III, \ref{AuIVE1}) Au IV, and \ref{HgVE1}) Hg V. The strongest transitions occur at shorter wavelengths, with most of them being shorter than 200 nm. We note that while the bulk of the transitions are at UV wavelengths, the density of lines may make this wavelength range difficult to use for line identification. Observing at longer wavelengths, where the lines are more sparse, may provide opportunities for more definitive identification of the charge state of the ion (see below). The weighted transition probabilities (gA, where g = $2J+1$), are compared with a HULLAC calculation \citep{tanaka2020systematic} and DESIRE \citep{fivet2007transition} for Os I \citep{quinet2006transition} and Ir II \citep{xu2007improved} in Table \ref{tabdesireosi} and \ref{tabdesireirii}, respectively. There is reasonable agreement between GRASP2K and DESIRE, and GRASP2K and HULLAC calculations. The average percent difference of Os I is $113\%$, and $126\%$, and the average percent difference of Ir II is $63.5\%$, and $82.7\%$, between GRASP2K and DESIRE, and GRASP2K and HULLAC calculations, respectively. In addition, the convergence of the E1 transition probability with different calculations for one transition is shown in Figure \ref{convergence}; it can be seen that the A-value is converged. The reason for choosing this transition is that for most of the ions in this paper, this transition has the strongest A-value, as shown in Table \ref{wavelength}. DESIRE does not provide this transition. Increasing the number of configurations was observed to have a minor impact on the transition rates of the most highly ionized systems in the Os-sequence. However, near-neutral systems (e.g. Ir II) show slower convergence with respect to increasing numbers of configurations. \\

\subsection{LTE Spectra and Expansion Opacities}
\begin{figure*}[tbp!]
\begin{adjustwidth}{-1.5cm}{}
	\centering
	\subfloat[Os I]{\includegraphics[width=11cm]{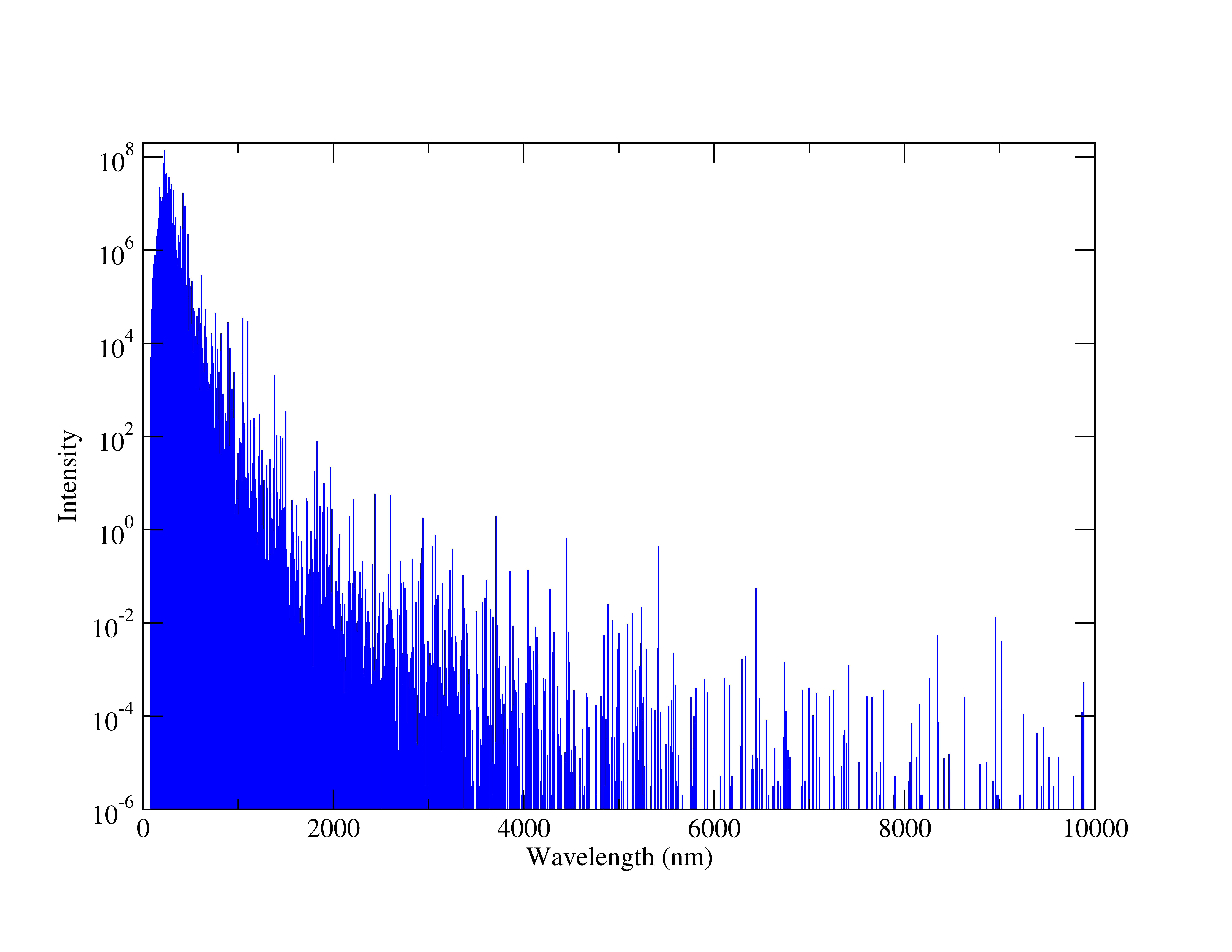}
		\label{OsIintensity}}
	\subfloat[Ir II]{\includegraphics[width=11cm]{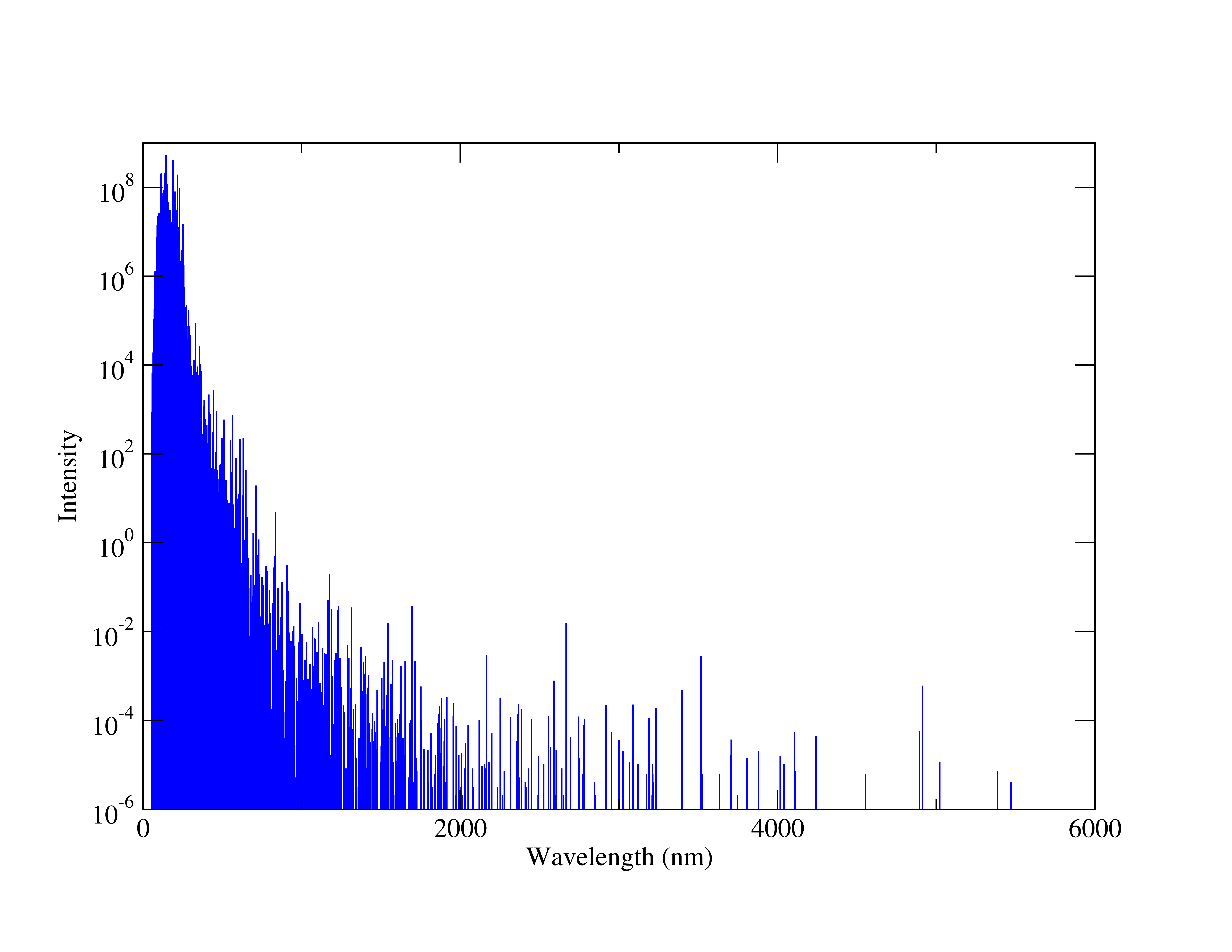}
		\label{IrIIintensity}}
	\hfill
	\subfloat[Pt III]{\includegraphics[width=11cm]{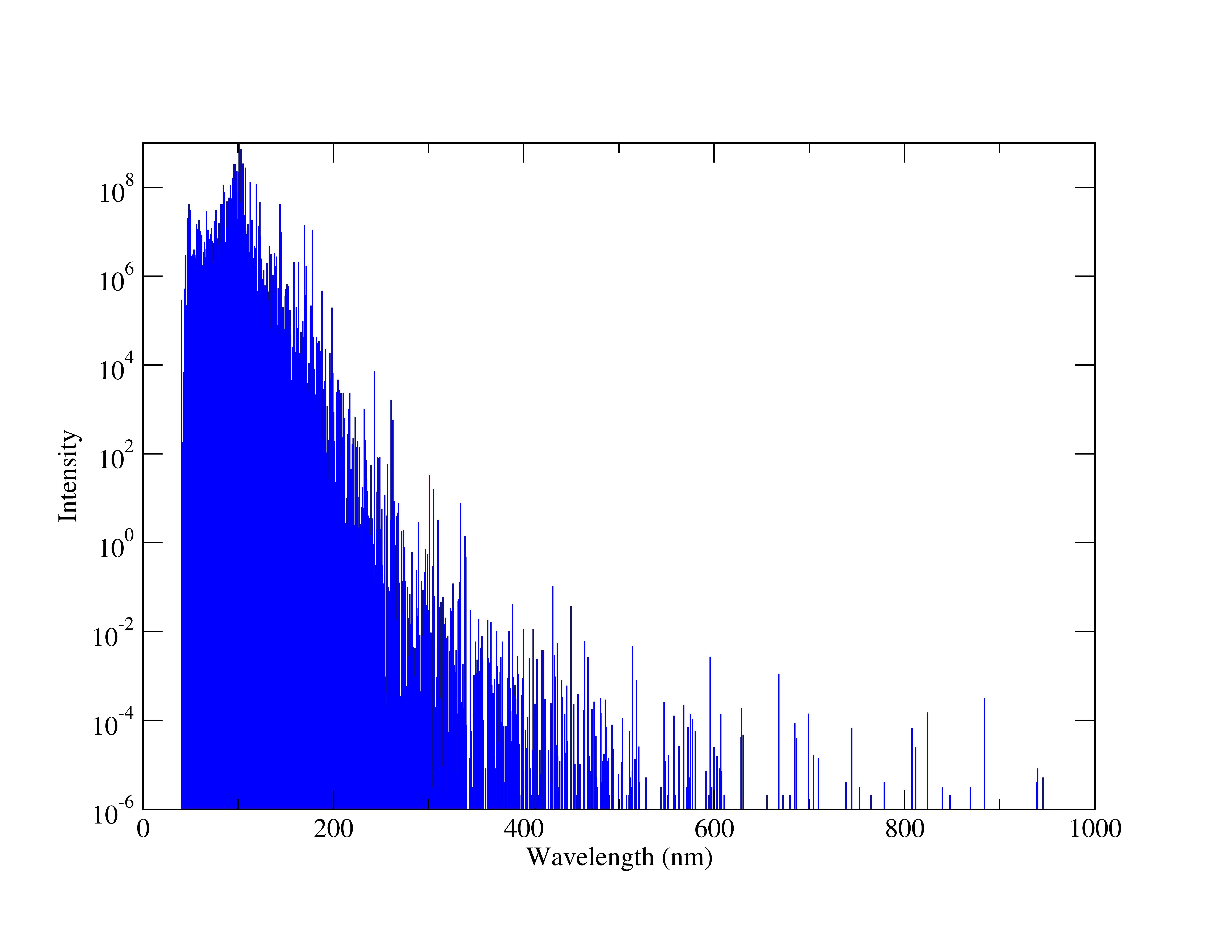}\label{PtIIIintensity}}	
	\subfloat[Au IV]{\includegraphics[width=11cm]{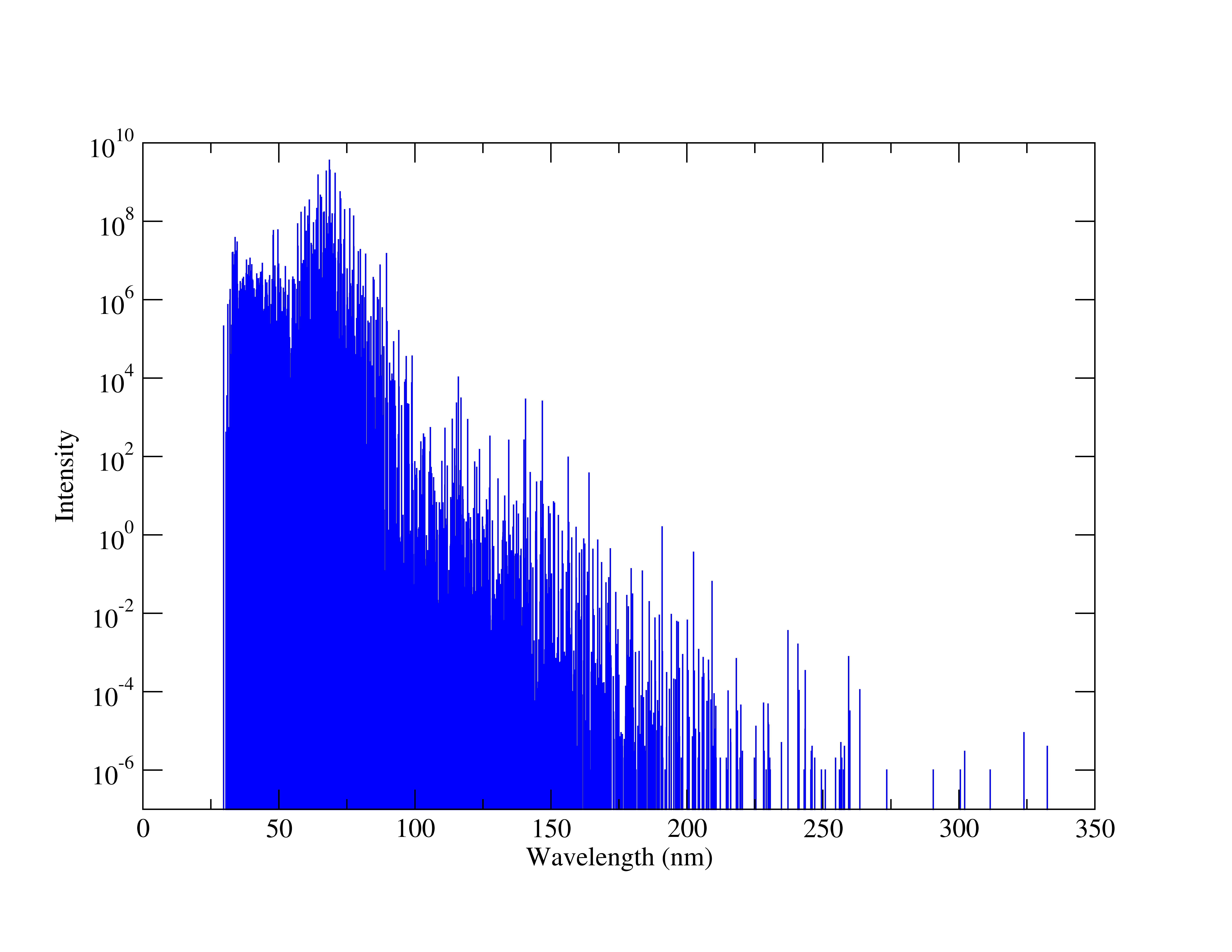}\label{AuIVintensity}}
		\hfill
	\hspace*{3cm}
	\subfloat[Hg V]{\includegraphics[width=11cm]{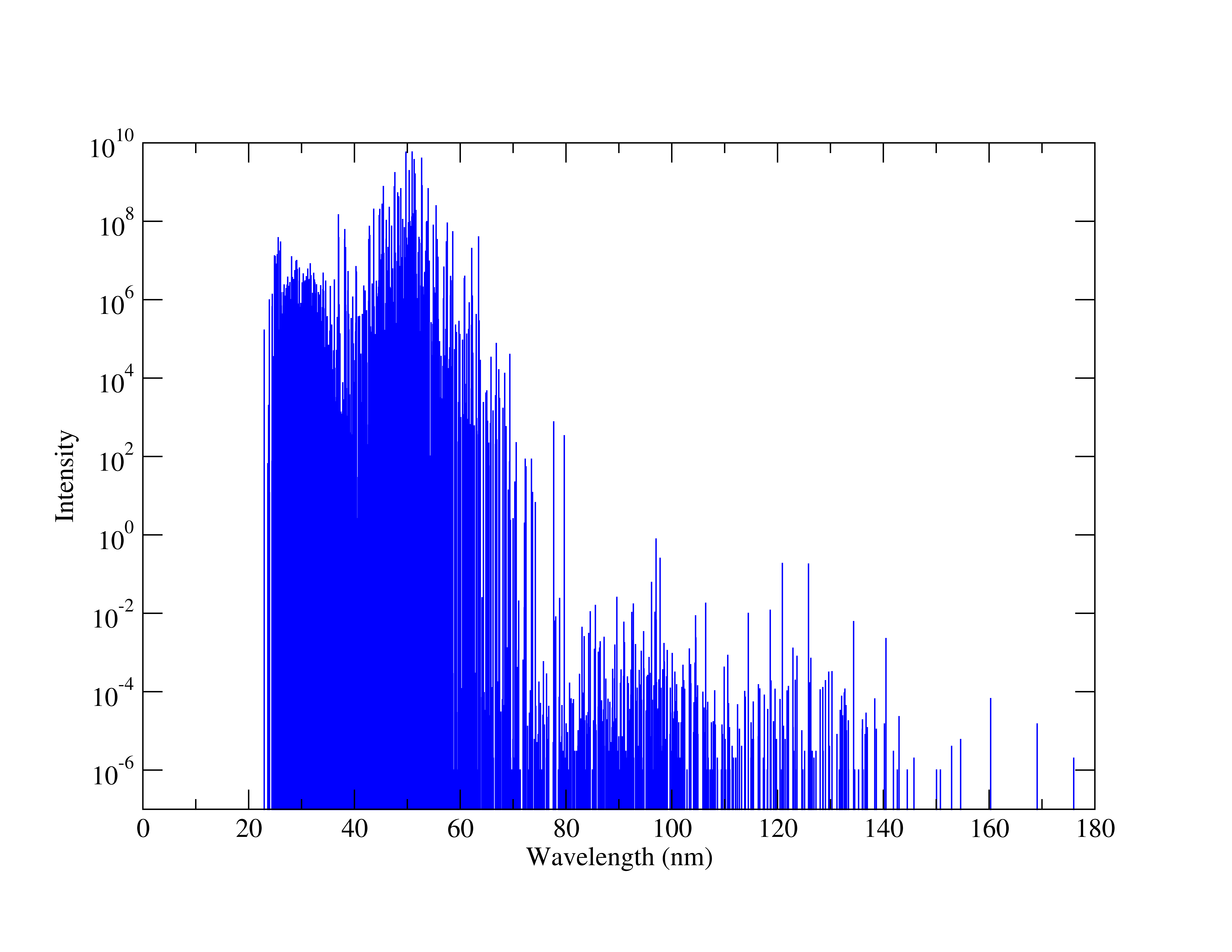}
		\label{HgVintensity}}
	\end{adjustwidth}
	\caption{LTE spectra at T=$5000$ K for a) Os I, b) Ir II, c) Pt III, d) Au IV, and e) Hg V.}
	\label{intensity}
\end{figure*}

The temperature in neutron star merger ejecta is approximately 5000~K, and cooling off to 1000~K at later times. Therefore, as an illustration, the LTE spectra for \ref{OsIintensity}) Os I, \ref{IrIIintensity}) Ir II, \ref{PtIIIintensity}) Pt III, \ref{AuIVintensity}) Au IV, and \ref{HgVintensity}) Hg V at 5000~K are computed and displayed for E1 transitions in ‌Figure \ref{intensity}. The emission lines are very dense in the UV and visible, but become less crowded as the wavelength increases into the IR. Also, wavelengths of the three most intense lines for each ion are represented in Table \ref{wavelength}.\\

\begin{figure*}[tbp!]
	\centering
	\subfloat{\includegraphics[width=15cm]{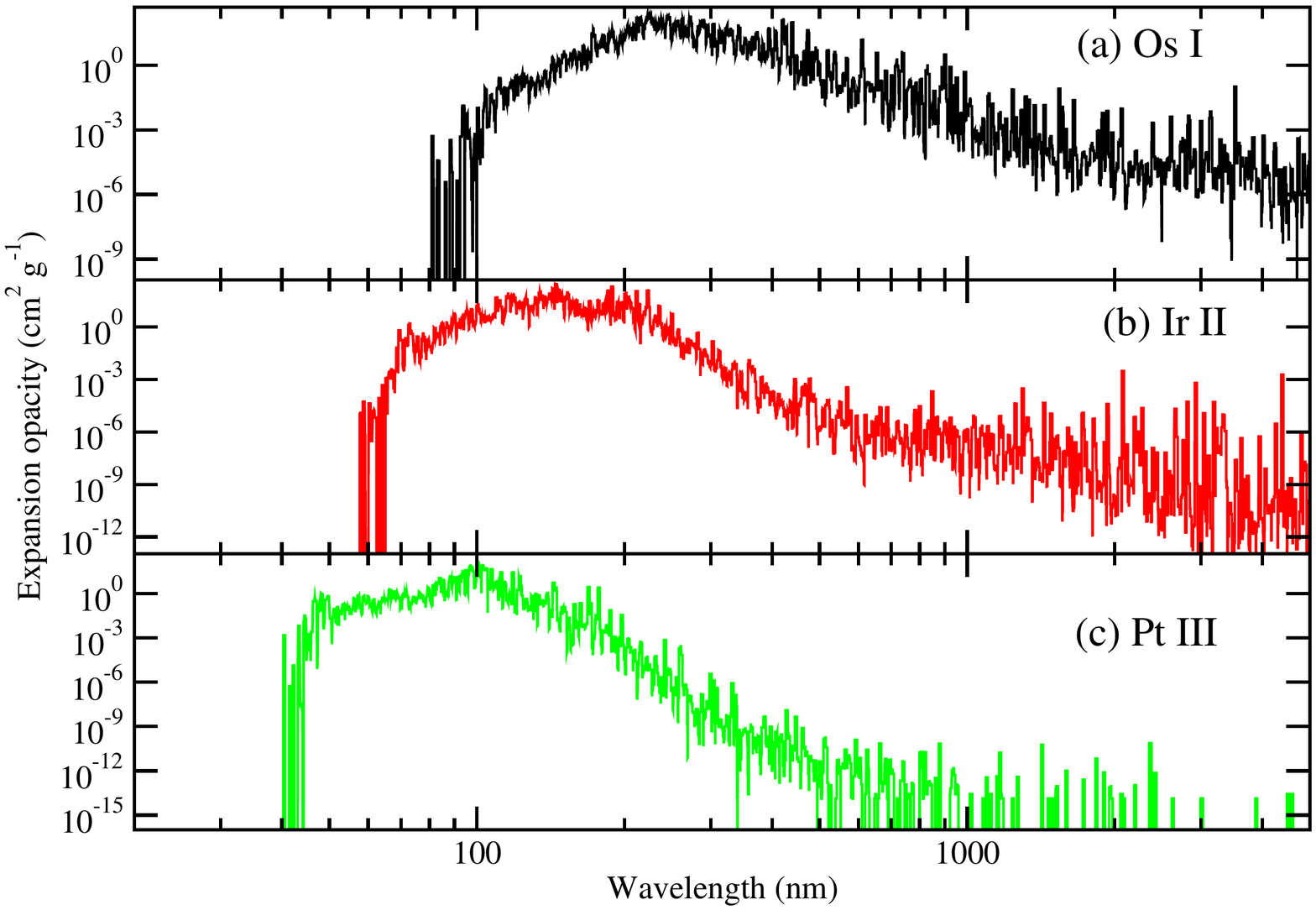}
		\label{OsIIrIIPtIIIopacity}}
	\hfill
	\subfloat{\includegraphics[width=15cm]{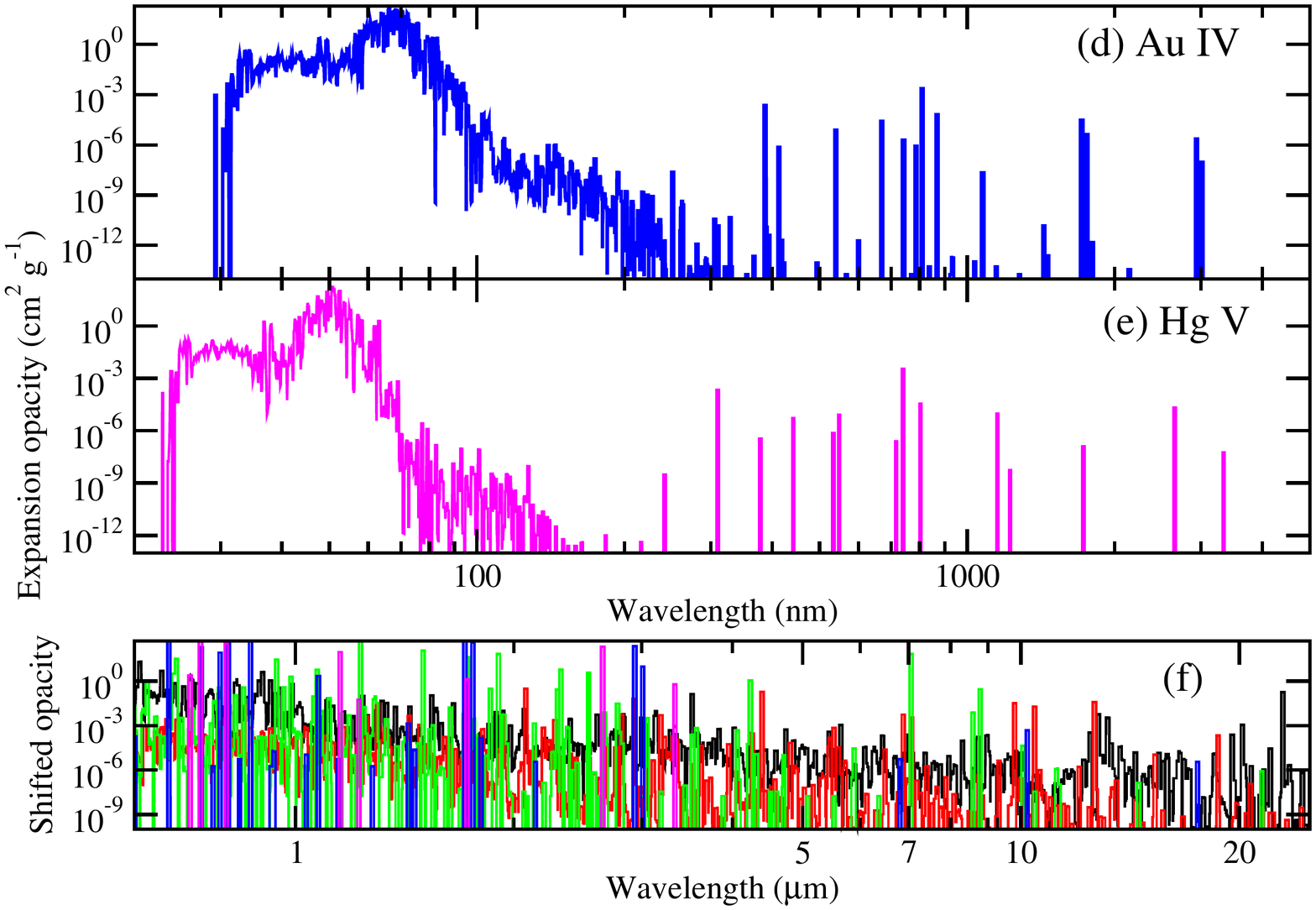}
		\label{AuIVHgVopacity}}
	\caption{Expansion opacity at 1 day for 3700 K and 10$^{-13}$ g/cm$^{3}$ for a) Os I, b) Ir II, c) Pt III, d) Au IV, and e) Hg V. The opacities in the IR are displayed in f) for all ions but with their magnitudes shifted for comparison.}
	\label{opacity}
\end{figure*}

In Figure \ref{opacity}, we plot the expansion opacity as given by Eq. \ref{eq:opacity} for each of the ions considered in this work. To facilitate comparison to prior studies \citep{kasen2013opacities, tanaka2020systematic}, the opacity is computed for $t=1$ day and $\rho=10^{-13}$ g/cm$^3$ with wavelength binning of $\Delta\lambda = 0.01 \lambda$. We choose a temperature of 3700~K as it is representative of a kilonove. The opacity peaks in the visible shifting into the UV with increasing ion charge from $\sim$225 nm for Os I to $\sim$50 nm for Hg V similar to the behavior of the LTE spectra in Figure \ref{intensity}. Further, as the ion charge increases, a number of IR lines appear with increasing strength over a less crowded
background. \\

Figure \ref{opacity}f provides a comparison of the IR lines within the wavelength window of the {\it James Webb Space Telescope (JWST)} for 5-25 $\mu$m. The background opacity magnitude in Figure \ref{opacity}f have been shifted to facilitate comparison of the lines for the various ions. A number of very prominent features emerge from the background and it is expected that in the mid-IR these features will be in emission.
Table \ref{wavelengthFS} lists the three most dominant transitions per ion which we predict may be observable by the Near Infrared Spectrograph (NIRSpec) or the Mid Infrared Instrument (MIRI) on {\it JWST} for a kilonova, particularly at late-times. However, there is some uncertainty in the
line positions as many transitions are between excited states for which there are practically no
experimental data; exceptions are for some lines due to Os I, Ir II, and Au IV as indicated in 
Table \ref{wavelengthFS}. The IR lines correspond primarily to M1 transitions within the ground configuration, and therefore the corresponding levels should be readily populated in the low temperature kilonova ejecta. \\

For Pt III, Gillanders et al. \citep{gillanders2021constraints} predict the $^3F_3 - ^3F_4$ line to occur at 1.092 $\mu$m, while the current work predicts the wavelength to be 1.234 $\mu$m. This aligns with a feature at 8.7 days in the observed spectrum of AT2017gfo. Further, a broad line centered near 1 $\mu$m may be explained by a cluster of Pt III lines at 0.947, 0.949, 0.991, and 1.077 $\mu$m (see Fig. 5f and Fig. 4 of Gillanders et al. \citep{gillanders2021constraints}). However, MIRI/{\it JWST} observations near 7 $\mu$m and 8.7 $\mu$m may be the best hope of identifying platinum in a kilonova.

\section{Summary}
In this work, we have performed atomic structure calculations for Os I (Z=76), Ir II (Z=77), Pt III (Z=78), Au IV (Z=79), and Hg V (Z=80) to construct atomic data for r-process elements. By using two different atomic code packages, GRASP2K and GRASP$^0$, energy levels,  and transition probabilities (E1 and M1) for the above ions were computed.

A list of the strongest transitions are given for each ion in the UV for E1 transitions and in the IR for M1 transitions. It is pointed out that the possible identification of r-process elements will be easier in the near- to mid-IR in the nebula phase of a kilonova and the M1 lines are far less crowded in this spectral window. The calculated data are made available in standard formats for use in the modeling of neutron star mergers at www.physast.uga.edu/ugamop/.
\\
\\

\renewcommand{\abstractname}{Acknowledgements}
\begin{abstract}
 This work was partially supported by NSF grant 1816984. 
	B M McLaughlin acknowledges the University of Georgia at 
	Athens for the award of an adjunct professorship and 
	Queen's University Belfast for a visiting research 
	fellowship (\textsc{vrf}). 
	This work was supported in part by resources and technical
	expertise from the UGA Georgia Advanced Computing Resource 
	Center (\textsc{gacrc}).
	The authors also gratefully acknowledge the Gauss Centre for 
	Supercomputing e.V. (www.gauss-centre.eu) 
	for funding this project by providing computing time 
	on the GCS Supercomputer \textsc{hazel hen} at \\
	H\"{o}chstleistungsrechenzentrum Stuttgart (www.hlrs.de). 
\end{abstract}
\bibliography{Draft}
\bibliographystyle{aa}
\end{multicols*}
\vspace*{-3cm}
\vspace{1cm}
\begin{table}[H]
	\centering
	\begin{adjustbox}{width=1\textwidth}
	\begin{tabular}{cccccc} 
		\hline\hline
		Ion/Atom    & Ground Config.& \multicolumn{1}{c}{GRASP2K}  \\ 
		\hline
		Os I & [Xe]$4f^{14}5d^66s^2$   &$5d^8$, $5d^7\{6s, 6p\}$, $5d^6\{6s^2, 6s6p\}$ \\
		Ir II & [Xe]$4f^{14}5d^76s$   &$5d^8$, $5d^7\{6s, 6p\}$, $5d^6\{6s^2, 6s6p\}$\\
		Pt III & [Xe]$4f^{14}5d^8$   &$5d^8$, $5d^7\{6s, 6p\}$, $5d^6\{6s^2, 6s6p\}$\\
		Au IV & [Xe]$4f^{14}5d^8$   &$5d^8$, $5d^7\{6s, 6p\}$, $5d^6\{6s^2, 6s6p\}$\\
		Hg V & [Xe]$4f^{14}5d^8$   &$5d^8$, $5d^7\{6s, 6p\}$, $5d^6\{6s^2, 6s6p\}$\\
		\hline\hline
		   \multicolumn{3}{c}{GRASP$^0$}  \\ 
		   \cline{1-3} 
		10-Config. & 12-Config.  & 16-Config. \\  
		\hline
  $5d^8$ & $5d^8$ & $5d^8$ \\
$5d^7\{6s, 6p, 6d\}$ & $5d^7\{6s, 6p, 6d\}$ &  $5d^7\{6s, 6p, 6d, 7s, 7p, 7d\}$ \\
$5d^6\{6s^2, 6p^2, 6d^2, 6s6p, 6s6d, 6p6d\}$ & $5d^6\{6s^2, 6p^2, 6d^2, 7s^2, 7p^2, 6s6p, 6s6d, 6p6d\}$ & $5d^6\{6s^2, 6p^2, 6d^2, 7s^2, 7p^2, 7d^2, 6s6p, 6s6d, 6p6d\}$ \\
		\hline\hline
	\end{tabular}
	\end{adjustbox}
	\caption{GRASP2K and GRASP$^0$ target model}\label{target}
\end{table}

\newpage

\begin{table*}[tbp!]
 \begin{adjustwidth}{-1.7cm}{}
	\begin{adjustbox}{width=1.2\textwidth}
	\begin{tabular}{ccccccccccccc}
		\hline\hline
		& Configuration & Term  & J     & NIST$^a$      &GRASP$^0$(12) &GRASP$^0$(15)       & GRASP2K(5)  &  HULLAC$^b(8)$ & $\Delta E^c$ & $\Delta E^d$ & $\Delta E^e$ & $\Delta E^f$\\
		\hline
		&$5d^66s^2$     & $^5D$ & 4     & 0.00          &0.00          &0.00                &0.00 & 0.00 & 0.00 & 0.00 & 0.00 & 0.00 \\
		&               &       & 2     & 2740.49       &3668.80       &3633.40             &3018.98 & 3669.99 & -928.31 & -278.49 & -1.19          & -651.01 \\
		&               &       & 3     & 4159.32       &3561.59       &3567.14             &4006.11 & 3485.71 & 597.73 & 153.21 & 75.88 &            520.4 \\
		&               &       & 1     & 5766.14       &5463.72       &5459.55             &5628.12 & 5384.29 & 302.42 & 138.02 & 79.43    & 243.83 \\
		&               &       & 0     & 6092.79       &5914.62       &5916.14             &6108.75 & 5846.54 & 178.17 & -15.96 & 68.08 &            262.21 \\
		\\
		&$5d^7(^4F)6s$  & $^5F$ & 5     & 5143.92       &7633.67       &6955.50             &2416.75 & 8297.05 & -2489.75 & 2727.17 & -663.38 & -5880.3 \\
		&               &       & 4     & 8742.83       &10897.65      &10189.09            &5683.66 & 11476.98 & -2154.82 & 3059.17 &        -579.33 & 5795.32 \\
		&               &       & 2     & 10165.98      &11856.53      &11599.45            &9282.52 & 14725.19 & -1690.55 & 883.46 &         -2868.66 & -5442.67 \\
		&               &       & 3     & 11378.00      &13234.12      &12590.30            &7963.28 & 13340.77 & -1856.12 & 3414.72                & -106.65 & -5377.49 \\
		&               &       & 1     & 13020.07      &15248.91      &14621.28            &10147.57 & 15321.69 & -2228.84 & 2872.50 & -72.78 & -5174.12 \\
		\\
		&$5d^7(^4F)6s$  & $^3F$ & 4     & 11030.58      &13462.15      &13181.68            &15942.40& 24083.44 & -2431.57 & -4911.82 & -10621.29 & -8141.04 \\
		&               &       & 2     & 12774.38      &14693.95      &14075.83            & 26417.68 & 21569.16 & -1919.57 & -13643.3 & -6875.21 & 4848.52 \\
		&               &       & 3     & 14091.37      &16657.69      &16183.66            &20061.03 & 21099.31 & -2566.32 & -5969.66 &     -4441.62 & -1038.28 \\
		\\
		&$5d^66s^2$     & $^3H$ & 5     & 14338.99      &16532.24      &16480.73            &18298.92 & 16145.83 & -2193.25 & -3959.93 & 386.41 & 2153.09 \\
		&               &       & 4     & 14848.05      &18223.93      &17245.52            & 21435.88 & 13304.98 & -3375.88 & -6587.83 & 4918.95 & 8130.9 \\
		&               &       & 6     & 14852.33      &16646.16      &16621.41            &15797.42 & 16351.83 & 1793.83 & -945.09 & 294.33 & -554.41 \\
		\\
		&$5d^7(^4P)6s$  & $^5P$ & 1     &      --      &19123.80      &18766.30            &16269.95 & 23138.85 & -- & -- & -4015.05 &       -6868.9 \\
		&               &       & 2     &    --         &16479.35      &16008.31            &16391.99 & 26552.53 & -- & -- & -10073.18              & -10160.54 \\
		&               &       & 3     & 15390.76      &20837.23      &20665.70            & 15827.03 & 27092.17 & -5446.47 & -436.27 & -6254.94 & -11265.14        \\
		\\
		&$5d^66s(^6D)6p$&$^7D^o$& 1     &         --    &18018.56      &18001.46            &13798.29 & 14018.25 & -- & -- & 4000.31 & -219.96 \\
		&               &       & 4     & 22615.69      &13418.13      &13403.97            &9079.96 & 9597.03 & 9197.65 & 13517.73 &  3821.10 & -517.07 \\
		&               &       & 5     & 23462.90      &13730.69      &13713.74            &9529.18 & 9941.73 & 9732.21 & 13933.72 & 3788.96 & -412.55 \\
		&               &       & 3     & 25012.93      &15506.13      &15491.82            &11228.11 & 11573.36 & 9506.80 & 13784.82 &       3932.77 & -345.25 \\
		&               &       & 2     & 25275.42      &16828.11      &16812.03            &12571.95 & 12853.23 & 8447.31 & 12703.47 & 3974.88 & -281.28 \\
		\\
		&$5d^66s(^6D)6p$&$^7P^o$& 2     &         --    &21136.37      &21120.13            &21650.15 & 23418.79 & -- & -- & -2282.42 & -1768.64 \\
		&               &       & 4     & 28331.77      &19935.86      &19919.65            & -- & 19087.33 & 8395.91 & -- & 848.53 & -- \\
		&               &       & 3     & 28371.68      &20829.29      &20812.61            &19925.91 & 21615.17 & 7542.39 & 8445.77         & -785.88 & -1689.26 \\
		\\
		&$5d^66s(^6D)6p$&$^7F^o$& 5     &  --           &20840.01      &20820.07            & 16297.80 & 17640.90 & -- & -- & 3199.11 & -1343.1 \\
		&               &       & 2     &    --         &28666.02      &26452.48            &16527.91 & 17633.15 & -- & -- & 11032.87               & -1105.24 \\
		&               &       & 4     &        --     &22321.80      &22302.33            &15195.33 & 16686.90 & -- & -- & 5634.9 & -1491.57 \\
		&               &       & 1     &      --       &21167.55      &21152.65            &16539.25 & 17589.88 & -- & -- & 3577.67 & -1050.63 \\
		&               &       & 3     &         --    &24695.57      &24674.47            &16189.48 & 17430.42 & -- & -- & 7265.15 & -1240.94 \\
		&               &       & 0     &         --    &21017.96      &21003.91            &16335.38 & 17416.14 & -- & -- & 3601.82 & -1080.76 \\
		&               &       & 6     &29099.41       &19162.43      &19140.88            &14586.66 & 16014.09 & 9936.98 & 14512.75 & 3148.34     & -1427.43 \\
		\dots\\
		\hline\hline
	\end{tabular}	
	\end{adjustbox}
	\caption{Energy levels in cm$^{-1}$ for Os I. The GRASP$^0$, GRASP2K, and HULLAC calculations have been performed with configuration numbers given in parentheses.}\label{tabosi}
	\raggedright 
	 \vspace{0.2cm}
	 $^a$ Atomic energy levels from the NIST database \citep{NIST_ASD, moore1958atomic}.\\
	 $^b$ Atomic energy levels from Tanaka et al. \citep{tanaka2020systematic} using the HULLAC code.\\
	 $^c$ Energy difference between NIST and GRASP$^0$(12). \\
         $^d$ Energy difference between NIST and GRASP2K. \\
	 $^e$ Energy difference between GRASP$^0$(12) and HULLAC. \\ 
         $^f$ Energy difference between GRASP2K and HULLAC. 
  \end{adjustwidth}
\end{table*}
\begin{table*}[tbp!]
	 \begin{adjustwidth}{-1.7cm}{}
	\begin{adjustbox}{width=1.2\textwidth}
	\small
	\begin{tabular}{cccccccccccccc}
		\hline\hline
		& Configuration & Term  & J     & NIST$^a$  & DESIRE$^b$     &GRASP$^0$(10) & GRASP$^0$(16) & GRASP2K(5) & HULLAC$^c(7)$ & $\Delta E^d$ & $\Delta E^e$ & $\Delta E^f$ & $\Delta E^g$\\
		\hline
		&$5d^7(^4F)6s$  & $^5F$ & 5     &  0.00     & 0.00        & 0.00               &0.00           &0.00 & 0.00 & 0.00 & 0.00 & 0.00 & 0.00 \\
		&               &       & 4     &  4787.93  & 4692.00     & 4392.63            &4297.37        &4544.01 & 4413.33 & 395.30 & 243.92 & -20.7 & 130.68 \\
		&               &       & 3     &  8186.96  & 8277.00     & 7578.85            &7540.14        &7740.31 & 7511.43 & 608.11 & 446.65 & 67.42 & 228.88 \\
		&               &       & 2     & 11307.32  &11374.00     & 8400.55            &9412.94        &10331.82 & 8731.69 & 2906.77 & 975.50 & -331.14 & 1600.13 \\
		&               &       & 1     & 11957.70  &12103.00     &10226.85            &10127.30       &10466.53 & 10173.43 & 1730.85 & 1511.17 & 53.42 & 293.10 \\
		\\
		&$5d^8$         & $^3F$ & 4     &  2262.75  & 2268.00     & 9786.89            &6657.87        &12345.86 & 13102.15 & -7524.14 & -1083.11 & -3315.26 & -756.29 \\
		&               &       & 3     &  9927.83  & 9838.00     &14900.59            &12920.44       &18062.12 & 18636.08 & -4972.76 & -8134.17 & -3735.49 & -573.96 \\
		&               &       & 2     & 17413.24  &17692.00     &9910.22            &14098.24       &22352.96 & 22390.63 & 7503.02 & -4939.72 &           -12480.41 & -37.67 \\	
		\\
		&$5d^8$         & $^3P$ & 2     &  3090.17  & 3266.00     &16399.17           &16057.24       & -- & 36459.05 & -13309.00 & -- & -20059.88 & -- \\
		&               &       & 1     &  9062.14  & 9014.00     &18553.06           &13368.85       &42438.61 & 42915.57 & -9490.92 & -33376.47 &        -24362.51 & -476.96 \\
		&               &       & 0     & 11211.93  &11134.00     &16812.57           &14685.56       & 40965.51 & 41500.78 & -5600.64 & -29753.58 & -24688.21 & -535.27 \\
		\\
		&$5d^8$         & $^1D$ & 2     &  8975.01  & 8867.00     &18023.33           &18327.27       & -- & 47691.11 & -9048.32 & -- & -29667.78 & -- \\
		\\
		&$5d^7(^4F)6s$  & $^3F$ & 4     & 11719.09  &11639.00     &14183.47           &10512.92       &15417.13 & 10181.02 & -2464.38 & -3698.04 & 4002.45           & 5236.11 \\
		&               &       & 3     & 17499.29  &17499.00     &15542.50           &15031.34       &20656.18 & 20745.07 & 1956.79 & -3156.89 & -5202.57         & -88.89 \\
		&               &       & 2     & 22467.78  &22351.00     &20903.68           &19787.66       &26129.89 & 19864.06 & 1564.10 & -3662.11 & 1039.62 & 6265.83 \\
		\\
		&$5d^7(^4P)6s$  & $^5P$ & 3     & 12714.64  &12808.00     &18881.64           &15462.13       &15719.57 & 15417.58 & -6167.00 & -3004.93 & 3464.06 & 301.99 \\
		&               &       & 2     & 15676.35  &15594.00     &21657.44           &22376.16       &18121.80 & 18348.06 & -5999.09 & -2445.45 & 3309.38 & -226.26 \\   
		&               &       & 1     & 18676.50  &18604.00     &18553.06           &17878.59       &22858.00 & 15037.46 & 123.44 & -4181.50 & 3515.6 & 7550.54 \\
		\\
		&$5d^8$         & $^1G$ & 4     & 17210.14  &17333.99     &15079.92           &13113.93       &37427.14 & 42637.17 & 2130.22 & 123.44 & -27557.25 & -5210.03 \\
		\\
		&$5d^7(^2G)6s$  & $^3G$ & 5     & 17477.92  &17440.00     &19332.94           &19154.40       &19729.27 & 19365.93 & -1855.02 & -2251.35 & -32.99 & 363.34 \\
		&               &       & 4     & 20294.23  &20317.00     &21701.91           &21012.21       & -- & 21113.67 & -1407.68 & -- & 588.24 & -- \\
		&               &       & 3     & 23195.21  &23122.00     &20305.01           &18229.64       &25481.87 & 24945.96 & 2890.20 & -2286.66 & -4640.95 & 535.91 \\
		\\
		&$5d^7(^4P)6s$  & $^3P$ & 2     & 18944.93  &18970.00     &25459.52           &25340.26       & -- & 26027.90 & -6550.59 & -- & -568.38 & -- \\
		&               &       & 1     & 20440.66  &20494.00     &23091.32           &21999.04       &25764.70 & 22409.70 & -3650.66 & -5324.04 & 681.62          & 3355.00 \\
		\\
		&$5d^66s^2$     & $^5D$ & 4     & 19279.05  &19217.00     &24103.48           &22690.11       &21725.60 & 15855.13 & -4824.43 & -2446.55 &           8248.35 & 5870.47 \\
		&               &       & 3     & 23727.67  &23681.00     &24841.36           &24421.62       &27514.10 & 26549.94 & -1113.69 & -3786.43 &          -1708.58 & 964.16 \\
		&               &       & 2     & 25563.72  &25484.00     &25459.52           &29233.36       &27861.34 & 30068.05 & 68.20 & -2297.62 & -4608.53 &    -2206.71 \\
		&               &       & 1     & 28600.35  &28725.00     &25531.45           &24137.33       &32703.21 & 31786.78 & 3068.90 & -4102.86 & -6255.33 & 916.43 \\
		\\	
		&$5d^7(^2H)6s$  & $^3H$ & 6     & 22267.00  &22436.00     &24404.96           &24200.01       &75768.56 & 24819.68 & -2137.96 & -53501.65 & -414.72 & 50948.88 \\
		&               &       & 5     & 25011.15  &24938.00     &26557.89           &26192.99       &77210.42 & 36343.71 & -1546.74 & -52199.27 &         -9785.82 & 40866.71 \\
		\\
		&$5d^7(^2D2)6s$ & $^3D$ & 3     & 26391.40  &26310.00     &27838.83           &27527.03       &28943.95 & 28333.41 & -1447.43 & -2552.55 & -494.58         & 610.54 \\
		\\
		&$5d^7(^2F)6s$  & $^3F$ & 3     & 31518.56  &31445.00     &35297.24           &29233.36       &38706.84 & 38230.60 & -3778.68 & -7188.28 & -2933.36 & 476.24 \\
		\\
		&$5d^7(^2G)6s$  & $^1G$ & 4     & 34319.48  & --           &28544.96           &26202.92       &25453.56 & 25167.13 & 5774.52 & 8865.92 & 3377.83 & 286.43 \\
		\\
		&$5d^7(^2P)6s$  & $^3P$ & 2     & 36160.59  & --           &27438.80           &32071.58       &43505.18 & 38790.16 & 8721.79 & -7344.59 & -11351.36 & 4715.02 \\
		\\
		&$5d^66s^2$     & $^3H$ & 5     & 36916.82  &36960.00     &32820.88           &32051.19       &42698.55 & 41484.66 & 4095.94 & -5781.73 & -8663.78 & 1213.89 \\
		\hline\hline
	\end{tabular}
	\end{adjustbox}	
	\caption{Energy levels in cm$^{-1}$ for Ir II.  The GRASP$^0$, GRASP2K, and HULLAC 
		calculations have been performed with configuration numbers given in parentheses.}
	\label{tabirii}
	\raggedright
	 \vspace{0.2cm} 
	 $^a$ Atomic energy levels, NIST database \citep{NIST_ASD, van1978term}.\\
	 $^b$ Atomic energy levels, DESIRE database \citep{fivet2007transition, xu2007improved}, Relativistic Hartree-Fock plus core polarization (HFR + CP). \\
	 $^c$Atomic energy levels from Tanaka et al. \citep{tanaka2020systematic} using the HULLAC code.\\
         $^d$ Energy difference between NIST and GRASP$^0$(10). \\
	 $^e$ Energy difference between NIST and GRASP2K. \\
         $^f$ Energy difference between GRASP$^0$(10) and HULLAC. \\ 
         $^g$ Energy difference between GRASP2K and HULLAC.
	\end{adjustwidth}
\end{table*}
\begin{table*}[h!]
	\centering
	
	\begin{tabular}{ccccccc} 
		\hline\hline
		\multicolumn{6}{c}{Os I}  \\ 
		 Configuration & Term  & J     & GRASP2K(4) & GRASP2K(5) & GRASP$^0$(12) & GRASP$^0$(15) \\ 
		 	\hline
		$5d^66s^2$     & $^5D$ & 4     & 0.00          &0.00           & 0.00                &0.00\\
		               &       & 2     & 3515.19 &3018.98      &3668.80       &3633.40              \\
		               &       & 3     &  3854.65 &4006.11      &3561.59       &3567.14              \\
		               &       & 1     & 5720.24  &5628.12     &5463.72       &5459.55               \\
		               &       & 0     & 6216.58  &6108.75     &5914.62       &5916.14              \\
		\hline\hline 
		\multicolumn{6}{c}{Ir II}  \\ 
		Configuration & Term  & J     & GRASP2K(4) & GRASP2K(5) & GRASP$^0$(10) & GRASP$^0$(16) \\  
		\hline
		$5d^7(^4F)6s$  & $^5F$ & 5     &  0.00            & 0.00               &0.00           &0.00  \\
		               &       & 4     &  4732.38  &4544.01     & 4392.63            &4297.37         \\
		               &       & 3     &  8037.62  &7740.31     & 7578.85            &7540.14        \\
		               &       & 2     & 9620.11   &10331.82    & 8400.55            &9412.94         \\
		               &       & 1     & 10855.25  &10466.53     &10226.85            &10127.30       \\
		\hline\hline 
		\multicolumn{6}{c}{Pt III}  \\ 
		Configuration & Term  & J     & GRASP2K(4) & GRASP2K(5) & GRASP$^0$(10) & GRASP$^0$(16) \\  
		\hline
		$5d^8$  & $^3F$ & 4     &  0.00            & 0.00               &0.00           &0.00  \\
		&       & 3     &  8388.75  & 8089.99     & 8436.73            &8618.15         \\
		&       & 2     &  14123.55  &13312.88     & 6330.84            &6609.90        \\
		\hline\hline 
		\multicolumn{6}{c}{Au IV}  \\ 
			Configuration & Term  & J     & GRASP2K(4) & GRASP2K(5) & GRASP$^0$(8) & GRASP$^0$(16) \\  
		\hline
		$5d^8$  & $^3F$ & 4    &  0.00            & 0.00               &0.00           &0.00  \\
		&       & 3     &  10945.77  &10689.7     & 11094.69            &11189.87         \\
		\hline\hline 
		\multicolumn{6}{c}{Hg V}  \\  
		Configuration & Term  & J     & GRASP2K(4) & GRASP2K(5) & GRASP$^0$(10) & GRASP$^0$(16) \\  
		\hline
	$5d^8$  & $^3F$ & 4    &  0.00            & 0.00               &0.00           &0.00  \\
	&       & 3     &  13677.30  &13440.74     & 13827.29            &11189.87         \\
		\hline\hline
	\end{tabular}
	\caption{Energy comparison of GRASP2K and GRASP$^0$ with different configurations as given in parentheses for the ground term.}\label{energycomparison}
\end{table*}
\begin{table*}[tbp!]
	\small
	\begin{tabular}{cccccccc}
		\hline\hline
		 Configuration & Term  & J     & Belfast group$^a$  & GRASP2K(5) & HULLAC$^b(5)$ & $\Delta E^c$ & $\Delta E^d$ \\
		\hline
		$5d^8$  & $^3F$ & 4     &  0.00     &   0.00     &        0.00        &     0.00      &  0.00  \\
		               &       & 3     &  9159.88  &   8089.99   &     8888.95        &   -1069.89     & -798.96  \\
		               &       & 2     &  14798.78  &   13312.88   &     14596.35       &    -1485.90    & -1283.47  \\
		\\
		$5d^8$  & $^1D$ & 2     &  6776.39     &   6680.22     &       6683.36         &     -96.17
      & -3.14  \\
		\\
		$5d^6 6s^2$  & $^5D$ & 4     &  79582.08     &   64561.61    &      --     &   -15020.47     & --  \\
		\hline\hline
		&&&&\multicolumn{2}{ c }{M1 A-value (s$^{-1}$)}\\ \cline{5-6}
		Lower Level & J & Upper Level & J & Belfast group$^a$  & GRASP2K(5) & $\%$$D_i$  \\
		\hline
		$5d^8\ ^3F$ & 4 & $5d^8\ ^3F$ & 3 & 19.30 & 13.38 & 36.23 \\
	        $5d^8\ ^1D$ & 2 & $5d^8\ ^3F$ & 2 & 8.19 & 4.80 & 52.19 \\
		\hline\hline
	\end{tabular}
	\caption{Energy levels in cm$^{-1}$ and M1 transition for Pt III.  The GRASP2K and HULLAC
		calculations have been performed with configuration numbers given in parentheses.}
	\label{tabptiii}
	\raggedright 
	  \vspace{0.2cm}
	  $^a$ The GRASP$^0$ calculation performed by \citet{gillanders2021constraints}.\\
	  $^b$ Atomic energy levels from Tanaka et al. \citep{tanaka2020systematic} using the HULLAC code.\\
	  $^c$ Energy difference between GRASP2K and Belfast group. \\
	  $^d$ Energy difference between GRASP2K and HULLAC. \\
	  $\%$$D_i$ percent difference of GRASP2K and Belfast group. 
\end{table*}
\begin{table*}	
	\begin{adjustbox}{width=1\textwidth}
	\begin{tabular}{cccccccccc}
		\hline\hline
		&&&&&\multicolumn{3}{ c }{gA(s$^{-1}$)}\\ \cline{6-8}
		& Lower Level & J & Upper Level & J & DESIRE$^a$  & GRASP2K & HULLAC$^b$ & $\%$$D_i$ & $\%$$D_j$ \\
		\hline
		& $5d^66s^2\ ^5D_4$ & 4 & $5d^66s(^6D)6p\ ^5F_5$ & 5 & 4.35E+08 & 4.41E+08 & 1.16E+08 & 1.37 & 116.70  \\
		& $5d^66s^2\ ^5D_4$ & 4 & $5d^66s(^6D)6p\ ^7D_5$ & 5 & 1.35E+07 & 8.96E+05 & 6.51E+08 & 175.10 & 199.45 \\
		& $5d^66s^2\ ^5D_4$ & 4 & $5d^66s(^4D)6p\ ^3F_4$ & 4 & 1.34E+08 & 1.97E+08 & 1.63E+08 & 38.07 & 18.89  \\
		& $5d^66s^2\ ^5D_4$ & 4 & $5d^66s(^4D)6p\ ^5D_4$ & 4 & 2.44E+08 & 6.58E+08 & 4.23E+08 & 91.80 & 43.48 \\
		& $5d^66s^2\ ^5D_4$ & 4 & $5d^66s(^6D)6p\ ^5D_4$ & 4 & 8.44E+07 & 2.31E+08 & 2.16E+08 & 92.96 & 6.71 \\
		& $5d^66s^2\ ^5D_4$ & 4 & $5d^66s(^6D)6p\ ^5F_4$ & 4 & 2.06E+07 & 6.49E+05 & 2.58E+07 & 187.78 & 190.18 \\
		& $5d^66s^2\ ^5D_4$ & 4 & $5d^66s(^6D)6p\ ^7F_4$ & 4 & 2.78E+08 & 1.41E+06 & 7.69E+07 & 197.98 & 192.80 \\
		& $5d^66s^2\ ^5D_4$ & 4 & $5d^66s(^6D)6p\ ^7D_4$ & 4 & 2.17E+07 & 7.60E+05 & 3.17E+04 & 186.46 & 183.98 \\
		& $5d^66s^2\ ^5D_4$ & 4 & $5d^7(^4F)6p\ ^3D_3$ & 3 & 3.97E+08 & 1.20E+06 & 2.96E+07 & 198.79 & 184.42 \\
		& $5d^66s^2\ ^5D_4$ & 4 & $5d^66s(^6D)6p\ ^5P_3$ & 3 & 1.38E+08 & 8.41E+07 & 9.97E+07 & 48.54 & 16.97 \\
		& $5d^66s^2\ ^5D_4$ & 4 & $5d^66s(^6D)6p\ ^7P_3$ & 3 & 4.55E+06 & 6.08E+05 & 3.41E+06 & 152.85 & 139.47 \\
		& $5d^66s^2\ ^5D_2$ & 2 & $5d^7(^4F)6p\ ^3D_3$ & 3 & 4.29E+08 & 2.76E+07 & 8.22E+07 & 175.82 & 99.45 \\
		& $5d^66s^2\ ^5D_2$ & 2 & $5d^66s(^6D)6p\ ^5P_3$ & 3 & 5.69E+07 & 1.39E+07 & 9.67E+07 & 121.47 & 149.73 \\
		& $5d^66s^2\ ^5D_3$ & 3 & $5d^7(^4F)6p\ ^3D_3$ & 3 & 3.35E+07 & 5.38E+06 & 2.99E+08 & 144.65 & 192.93 \\
		& $5d^66s^2\ ^5D_3$ & 3 & $5d^66s(^6D)6p\ ^5P_3$ & 3 & 5.39E+06 & 5.99E+07 & 4.80E+07 & 166.98 & 22.06 \\
		& $5d^66s^2\ ^5D_3$ & 3 & $5d^66s(^6D)6p\ ^7P_3$ & 3 & 3.85E+06 & 6.41E+05 & 1.55E+08 & 142.91 & 198.35 \\
		& $5d^66s^2\ ^5D_3$ & 3 & $5d^66s(^4D)6p\ ^3F_4$ & 4 & 8.05E+07 & 3.11E+07 & 1.04E+07 & 88.53 & 99.76 \\
		& $5d^66s^2\ ^5D_3$ & 3 & $5d^66s(^4D)6p\ ^5D_4$ & 4 & 3.99E+07 & 7.36E+06 & 2.76E+06 & 137.71 & 90.91  \\
		& $5d^66s^2\ ^5D_3$ & 3 & $5d^66s(^6D)6p\ ^5D_4$ & 4 & 8.01E+06 & 3.93E+06 & 4.66E+06 & 68.34 & 16.70 \\
		& $5d^66s^2\ ^5D_3$ & 3 & $5d^66s(^6D)6p\ ^5F_4$ & 4 & 7.20E+07 & 1.80E+08 & 1.46E+08 & 85.71 & 20.86 \\
		& $5d^66s^2\ ^5D_3$ & 3 & $5d^66s(^6D)6p\ ^7F_4$ & 4 & 1.96E+07 & 2.27E+06 & 1.04E+07 & 158.48 & 128.33 \\
		& $5d^66s^2\ ^5D_3$ & 3 & $5d^66s(^6D)6p\ ^7D_4$ & 4 & 9.00E+04 & 1.82E+03 & 4.65E+05 & 192.07 & 198.44 \\
		& $5d^7(^4F)6s\ ^5F_4$ & 4 & $5d^7(^4F)6p\ ^3D_3$ & 3 & 3.65E+07 & 7.47E+07 & 9.45E+07 & 68.70 & 23.40 \\
		& $5d^7(^4F)6s\ ^5F_4$ & 4 & $5d^66s(^6D)6p\ ^5P_3$ & 3 & 1.18E+07 & 4.81E+05 & 7.17E+05 & 184.33 & 39.40 \\
		& $5d^7(^4F)6s\ ^5F_4$ & 4 & $5d^66s(^6D)6p ^7P_3$ & 3 & 9.80E+05 & 6.49E+03 & 2.31E+07 & 197.37 & 199.89 \\
		& $5d^7(^4F)6s\ ^5F_4$ & 4 & $5d^66s(^4D)6p ^3F_4$ & 4 & 2.44E+07 & 4.63E+07 & 5.51E+07 & 61.95 & 17.36 \\
		& $5d^7(^4F)6s\ ^5F_4$ & 4 & $5d^66s(^4D)6p ^5D_4$ & 4 & 2.48E+07 & 2.56E+07 & 3.17E+05 & 3.17 & 195.11 \\
		& $5d^7(^4F)6s\ ^5F_4$ & 4 & $5d^66s(^6D)6p ^5D_4$ & 4 & 5.62E+07 & 7.59E+07 & 7.99E+06 & 29.83 & 161.90 \\
		& $5d^7(^4F)6s\ ^5F_4$ & 4 & $5d^66s(^6D)6p ^5F_4$ & 4 & 6.48E+06 & 3.45E+06 & 1.72E+05 & 61.03 & 181.00 \\
		& $5d^7(^4F)6s\ ^5F_4$ & 4 & $5d^66s(^6D)6p ^7F_4$ & 4 & 9.00E+04 & 8.75E+04 & 5.45E+06 & 2.82 & 193.68 \\
		& $5d^7(^4F)6s\ ^3F_4$ & 4 & $5d^7(^4F)6p ^3D_3$ & 3 & 1.31E+08 & 2.03E+08 & 4.869E+07 & 43.11 & 122.62 \\
		& $5d^7(^4F)6s\ ^3F_4$ & 4 & $5d^66s(^6D)6p ^5F_4$ & 4 & 4.50E+05 & 3.22E+04 & 2.97E+05 & 173.29 & 160.87 \\
		& $5d^7(^4F)6s\ ^3F_4$ & 4 & $5d^66s(^6D)6p ^5F_5$ & 5 & 1.32E+06 & 1.43E+04 & 1.57E+06 & 195.71 & 196.39 \\
		& $5d^7(^4F)6s ^5F_3$ & 3 & $5d^7(^4F)6p ^3D_3$ & 3 & 4.96E+07 & 7.12E+07 & 2.05E+07 & 35.76 & 110.58 \\
		& $5d^7(^4F)6s ^5F_3$ & 3 & $5d^66s(^6D)6p ^5P_3$ & 3 & 3.64E+06 & 2.48E+06 & 2.73E+07 & 37.91 & 166.69 \\
		& $5d^7(^4F)6s ^5F_3$ & 3 & $5d^66s(^6D)6p ^7P_3$ & 3 & 6.30E+05 & 4.50E+04 & 4.64E+05 & 173.33 & 164.64 \\
		& $5d^7(^4F)6s ^5F_3$ & 3 & $5d^66s(^4D)6p ^3F_4$ & 4 & 9.54E+06 & 7.02E+06 & 2.47E+06 & 30.43 & 95.89 \\
		& $5d^7(^4F)6s ^5F_3$ & 3 & $5d^66s(^4D)6p ^5D_4$ & 4 & 1.26E+06 & 2.00E+07 & 1.17E+04 & 176.29 & 199.77 \\
		& $5d^7(^4F)6s ^5F_3$ & 3 & $5d^66s(^6D)6p ^5D_4$ & 4 & 5.40E+06 & 9.73E+05 & 5.61E+06 & 138.93 & 140.88 \\
		& $5d^7(^4F)6s ^5F_3$ & 3 & $5d^66s(^6D)6p ^5F_4$ & 4 & 2.70E+05 & 1.84E+05 & 2.50E+06 & 37.88 & 172.58 \\
		\dots\\
		\hline\hline
	\end{tabular} 
	\end{adjustbox}
	\caption{Weighed transition probabilities (gA-values) for Os I}\label{tabdesireosi}
	\raggedright 
        \vspace{0.2cm}
	 $^a$ gA-values, DESIRE database \citep{fivet2007transition, quinet2006transition}. \\
	 $^b$ gA-values from Tanaka et al \citep{tanaka2020systematic} using HULLAC.\\
	 $\%$$D_i$ percent difference of GRASP2K and DESIRE. \\
	 $\%$$D_j$ percent difference of GRASP2K and HULLAC.
\end{table*}
\begin{table*}	
	\begin{adjustbox}{width=1\textwidth}
	\begin{tabular}{cccccccccc}
		\hline\hline
		&&&&&\multicolumn{3}{ c }{gA(s$^{-1}$)}\\ \cline{6-8}
		& Lower Level & J & Upper Level & J & DESIRE$^a$  & GRASP2K & HULLAC$^b$& $\%$$D_i$ & $\%$$D_j$ \\
		\hline
		& $5d^7(^4F)6s\ ^5F$ & 5 & $5d^7(^4F)6p$ & 4 & 2.85E+08   & 2.55E+08 & 1.75E+08 & 11.11 & 37.21 \\
		& $5d^7(^4F)6s\ ^5F$ & 5 & $5d^7(^4F)6p$ & 4 & 5.10E+07   & 6.81E+07 & 4.80E+07 & 28.71 & 34.62 \\
		& $5d^7(^4F)6s\ ^5F$ & 5 & $5d^6(^5D)6s6p$ & 4 & 1.28E+08  & 9.03E+07 & 1.96E+08 & 34.54 & 73.84 \\
		& $5d^7(^4F)6s\ ^5F$ & 5 & $5d^7(^4F)6p$ & 4 & 1.60E+09  & 1.46E+09 & 1.37E+09 & 9.15 & 6.21 \\
		& $5d^7(^4F)6s\ ^5F$ & 5 & $5d^7(^4F)6p$ & 5 & 1.33E+09  & 2.93E+09 & 7.31E+08 & 75.12 & 120.13 \\
		& $5d^7(^4F)6s\ ^5F$ & 5 & $5d^6(^5D)6s6p$ & 5 & 1.28E+08  & 7.82E+08 & 2.23E+08 & 143.74 & 111.24 \\
		& $5d^7(^4F)6s\ ^5F$ & 5 & $5d^7(^4F)6p$ & 5 & 1.90E+09  & 2.93E+09 &  2.38E+09 & 42.65 & 20.72 \\
		& $5d^8\ ^3F$ & 4 & $5d^7(^4F)6p$ & 3 & 1.49E+07  & 5.76E+07 & 7.40E+07 & 117.79 & 24.92 \\
		& $5d^8\ ^3F$ & 4 & $5d^7(^4F)6p$ & 3 & 1.17E+07  & 9.00E+07 & 1.67E+08 & 153.98 & 59.92 \\
		& $5d^8\ ^3F$ & 4 & $5d^7(^4F)6p$ & 4 & 1.34E+08  & 7.94E+08 & 6.58E+07 & 142.24 & 169.39\\
		& $5d^8\ ^3F$ & 4 & $5d^6(^5D)6s6p$ & 4 & 3.80E+06  & 6.60E+07 & 6.09E+06 & 178.22 & 166.21 \\
		& $5d^8\ ^3F$ & 4 & $5d^7(^4F)6p$ & 5 & 2.03E+08  & 3.01E+08 & 6.99E+08 & 38.89 & 79.60 \\
		& $5d^8\ ^3F$ & 4 & $5d^7(^4F)6p$ & 5 & 1.17E+07 & 2.55E+07 & 8.03E+07 & 74.19 & 103.59 \\
		& $5d^7(^4F)6s\ ^5F$ & 4 & $5d^7(^4F)6p$ & 5 & 1.78E+09 & 1.22E+09 & 1.65E+09 & 37.33 & 29.96 \\
		& $5d^7(^4F)6s\ ^5F$ & 4 & $5d^6(^5D)6s6p$ & 5 & 4.50E+06 & 5.11E+08 & 3.01E+04 & 196.51 & 199.98 \\
		& $5d^7(^4F)6s\ ^5F$ & 4 & $5d^7(^4F)6p$ & 4 & 2.49E+08 & 1.27E+08 & 2.35E+07 & 64.89 & 137.54 \\
		& $5d^7(^4F)6s\ ^5F$ & 4 & $5d^7(^4F)6p$ & 4 & 1.59E+09 & 1.70E+09 & 2.21E+09 & 6.69 & 26.09 \\
		& $5d^7(^4F)6s\ ^5F$ & 4 & $5d^6(^5D)6s6p$ & 4 & 6.63E+07 & 5.86E+07 & 1.78E+08 & 12.33 & 100.93 \\
		& $5d^7(^4F)6s\ ^5F$ & 4 & $5d^7(^4F)6p$ & 4 & 1.48E+08 & 1.02E+08 & 8.01E+07 & 36.80 & 24.05 \\
		& $5d^7(^4F)6s\ ^5F$ & 4 & $5d^7(^4F)6p$ & 3 & 3.32E+07 & 3.98E+07 & 1.96E+06 & 18.08 & 181.23 \\
		& $5d^7(^4F)6s\ ^5F$ & 4 & $5d^6(^5D)6s6p$ & 3 & 1.81E+08 & 3.85E+08 & 2.70E+08 & 72.08 & 35.11 \\
		& $5d^7(^4F)6s\ ^5F$ & 4 & $5d^7(^4F)6p$ & 3 & 1.08E+09 & 9.78E+08 & 4.50E+07 & 9.91 & 182.40 \\
		& $5d^7(^4F)6s\ ^5F$ & 3 & $5d^7(^4F)6p$ & 4 & 1.89E+08 & 2.08E+08 & 4.03E+08 & 9.57 & 63.83 \\
		& $5d^7(^4F)6s\ ^5F$ & 3 & $5d^7(^4F)6p$ & 4 & 2.37E+08 & 1.24E+08 & 2.50E+08 & 62.60 & 67.38 \\
		& $5d^7(^4F)6s\ ^5F$ & 3 & $5d^6(^5D)6s6p$ & 4 & 5.00E+06 & 3.54E+07 & 2.30E+08 & 150.49 & 146.65 \\
		& $5d^7(^4F)6s\ ^5F$ & 3 & $5d^7(^4F)6p$ & 3 & 5.98E+07 & 2.17E+07 & 1.90E+07 & 93.50 & 13.27 \\
		& $5d^7(^4F)6s\ ^5F$ & 3 & $5d^6(^5D)6s6p$ & 3 & 1.98E+08 & 1.77E+08 & 8.06E+07 & 11.20 & 74.84  \\
		& $5d^7(^4F)6s\ ^5F$ & 3 & $5d^7(^4F)6p$ & 3 & 4.52E+08 & 1.15E+08 & 1.15E+08 & 118.87 & 0.00 \\
		& $5d^7(^4F)6s\ ^5F$ & 3 & $5d^7(^4F)6p$ & 2 & 1.17E+08 & 1.94E+08 & 2.10E+08 & 49.52 & 7.92 \\
		& $5d^7(^4F)6s\ ^5F$ & 3 & $5d^6(^5D)6s6p$ & 2 & 2.00E+07 & 3.89E+07 & 9.46E+06 & 64.18 & 121.75 \\
		& $5d^7(^4F)6s\ ^5F$ & 3 & $5d^7(^4F)6p$ & 2 & 3.61E+08 & 3.38E+08 & 1.82E+08 & 6.58 & 60.00 \\
		& $5d^7(^4F)6s\ ^5F$ & 3 & $5d^7(^4F)6p$ & 2 & 2.19E+08 & 5.02E+08 & 6.23E+08 & 78.50 & 21.51  \\
		& $5d^7(^4F)6s\ ^5F$ & 2 & $5d^7(^4F)6p$ & 3 & 2.00E+08 & 3.36E+08 & 9.88E+07 & 50.75 & 109.11 \\
		& $5d^7(^4F)6s\ ^5F$ & 2 & $5d^6(^5D)6s6p$ & 3 & 8.50E+06 & 7.84E+06 & 1.04E+07 & 8.08 & 28.07  \\
		& $5d^7(^4F)6s\ ^5F$ & 2 & $5d^7(^4F)6p$ & 3 & 5.56E+07 & 1.29E+07 & 1.67E+09 & 124.67 & 196.93 \\
		& $5d^7(^4F)6s\ ^5F$ & 2 & $5d^7(^4F)6p$ & 2 & 1.48E+08 & 1.87E+08 & 1.63E+08 & 23.28 & 13.71 \\
		& $5d^7(^4F)6s\ ^5F$ & 2 & $5d^6(^5D)6s6p$ & 2 & 1.16E+08 & 6.63E+08 & 9.31E+06 & 140.44 & 194.46 \\
		& $5d^7(^4F)6s\ ^5F$ & 2 & $5d^7(^4F)6p$ & 2 & 5.73E+07 & 9.87E+07 & 1.84E+04 & 53.08 & 199.92  \\
		& $5d^7(^4F)6s\ ^5F$ & 2 & $5d^7(^4F)6p$ & 2 & 1.22E+08 & 2.13E+08 & 2.91E+08 & 54.33 & 30.95 \\
		& $5d^7(^2P)6s\ ^3P$ & 1 & $5d^7(^4F)6p$ & 2 & 2.42E+08 & 2.18E+08 & 3.28E+08 & 10.43 & 40.29 \\
		& $5d^7(^2P)6s\ ^3P$ & 1 & $5d^7(^4F)6p$ & 2 & 2.98E+07 & 2.10E+07 & 5.26E+06 & 34.65 & 119.88 \\
		& $5d^7(^2P)6s\ ^3P$ & 1 & $5d^7(^4F)6p$ & 1 & 2.65E+07 & 1.33E+07 & 1.23E+07 & 66.33 & 7.81\\
		& $5d^7(^2P)6s\ ^3P$ & 1 & $5d^7(^4F)6p$ & 1 & 8.55E+07 & 1.05E+08 & 6.92E+07 & 20.47 & 41.10 \\
		& $5d^7(^2P)6s\ ^3P$ & 1 & $5d^7(^4F)6p$ & 0 & 1.47E+08 & 2.61E+08 & 3.16E+07  & 55.88 & 156.80  \\
		\dots\\
		\hline\hline
	\end{tabular}
	\end{adjustbox}
	\caption{Weighed transition probabilities (gA-values) for Ir II}\label{tabdesireirii}
	\raggedright 
	 \vspace{0.2cm}
	$^a$ gA-values, DESIRE database \citep{fivet2007transition, xu2007improved}. \\
	$^b$ gA-values from Tanaka et al \citep{tanaka2020systematic} using HULLAC.\\
	$\%$$D_i$ percent difference of GRASP2K and DESIRE. \\
	$\%$$D_J$ percent difference of GRASP2K and HULLAC.
\end{table*}
\begin{table*}[h!]
	\centering
	\begin{tabular}{ccc} 
		\hline\hline
		 \multicolumn{3}{c}{Os I}  \\ 
		Lower Level & Upper Level & $\lambda$(nm) \\ 
		\hline
		 $5d^6(^5D_4)6s^2\ ^5D$ &\  $5d^6(^5D_4)6s(^4D)6p\ ^5F$ &\ 226.43\\
		 $5d^6(^5D_4)6s^2\ ^5D$ &\  $5d^7(^4F_3)6p\ ^5G$ &\ 213.48\\
		 $5d^6(^5D_4)6s^2\ ^5D$ &\  $5d^6(^3H_4)6s(^2H)6p\ ^1G$&\ 224.99\\
		\hline\hline 
		  \multicolumn{3}{c}{Ir II}  \\ 
		  Lower Level & Upper Level & $\lambda$(nm) \\  
		 \hline
		 $5d^7(^4F_3)6s\ ^5F$ &\ $5d^6(^3H_4)6s(^4H)6p\ ^5G$&\ 145.93\\
		 $5d^7(^4F_3)6s\ ^5F$ &\ $5d^7(^4F_3)6p\ ^5G$&\ 189.69\\
		 $5d^7(^4F_3)6s\ ^5F$ & \ $5d^6(^5D_4)6s(^4D)6p\ ^5F$ &\ 115.80\\
		 \hline\hline 
		 \multicolumn{3}{c}{Pt III}  \\ 
		  Lower Level & Upper Level & $\lambda$(nm) \\  
		 \hline
		 $5d^8\ ^3F_2$ & \ $5d^7(^2H_3)6p\ ^3G$ &\ 101.08\\
		 $5d^7(^4F_3)6s\ ^5F$ &\ $5d^7(^2F_3)6p\ ^3F$ &\ 103.16\\
		 $5d^8\ ^3F_2$ &\ $5d^7(^2G_3)6p\ ^3H$&\ 105.14\\
		 \hline\hline 
		 \multicolumn{3}{c}{Au IV}  \\ 
		  Lower Level & Upper Level & $\lambda$(nm) \\  
		 \hline
		 $5d^8\ ^3F_2$ & \ $5d^7(^2H_3)6p\ ^3G$ &\ 68.59\\
		 $5d^8\ ^3F_2$ &\ $5d^7(^4F_3)6p\ ^3G$ &\ 68.80\\
		 $5d^8\ ^3F_2$ &\ $5d^7(^4F_3)6p\ ^3F$&\ 67.41\\
		 \hline\hline 
		 \multicolumn{3}{c}{Hg V}  \\  
		  Lower Level & Upper Level & $\lambda$(nm) \\  
		 \hline
		 $5d^8\ ^3F_2$ &\ $5d^7(^2H_3)6p\ ^3G$ &\ 50.36\\
		 $5d^7(^2F_3)6s\ ^3F$ &\ $5d^6(^5D_4)6s(^4D)6p\ ^3D$&\ 49.73\\
		 $5d^8\ ^3F_2$ &\ $5d^7(^4F_3)6p\ ^5F$ &\ 52.71\\
		\hline\hline
	\end{tabular}
	\caption{Wavelength of the most intense lines from LTE spectra at 5000 K.}\label{wavelength}
\end{table*}
\begin{table*}[h!]
	\centering
	\begin{tabular}{ccc} 
		\hline\hline
		 \multicolumn{3}{c}{Os I}  \\ 
		Lower Level & Upper Level & $\lambda$($\mu$m) \\ 
		\hline
		 $5d^66s^2\ ^5D_4$ &\  $5d^76s\ ^5F_5$ &\ 1.9440*\\
		 $5d^66s^2\ ^5D_2$ &\  $5d^66s6p\ ^7P_3$ &\ 3.540\\
		 $5d^76s\ ^5F_1$ &\  $5d^66s6p\ ^7D_1$&\ 12.85\\
		\hline\hline 
		  \multicolumn{3}{c}{Ir II}  \\ 
		  Lower Level & Upper Level & $\lambda$($\mu$m) \\  
		 \hline
		 $5d^76s\ ^5F_5$ &\ $5d^76s\ ^5F_4$&\ 2.0886*\\
		 $5d^76s\ ^5F_5$ &\ $5d^8\ ^3F_4$&\ 4.4201*\\
		 $5d^76s\ ^5F_3$ & \ $5d^66s6p\ ^5F_2$ &\ 12.689*\\
		 \hline\hline 
		 \multicolumn{3}{c}{Pt III}  \\ 
		  Lower Level & Upper Level & $\lambda$($\mu$m) \\  
		 \hline
		 $5d^8\ ^3F_4$ & \ $5d^8\ ^3F_3$ &\ 1.234\\
		 $5d^76s\ ^5F_3$ &\ $5d^76s\ ^5F_2$ &\ 4.257\\
		 $5d^8\ ^1D_2$ &\ $5d^8\ ^3F_3$&\ 7.09\\
		 \hline\hline 
		 \multicolumn{3}{c}{Au IV}  \\ 
		  Lower Level & Upper Level & $\lambda$($\mu$m) \\  
		 \hline
		 $5d^8\ ^3F_4$ & \ $5d^8\ ^3F_3$ &\ 0.81344*\\
		 $5d^8\ ^3F_3$ &\ $5d^8\ ^3P_2$ &\ 1.7218*\\
		 $5d^8\ ^3P_0$ &\ $5d^8\ ^3P_1$&\ 3.0261*\\
		 \hline\hline 
		 \multicolumn{3}{c}{Hg V}  \\  
		  Lower Level & Upper Level & $\lambda$($\mu$m) \\  
		 \hline
		 $5d^8\ ^3F_4$ &\ $5d^8\ ^3F_3$ &\ 0.744\\
		 $5d^8\ ^1D_2$ &\ $5d^8\ ^3F_3$&\ 2.665\\
		 $5d^8\ ^3P_0$ &\ $5d^8\ ^3P_1$ &\ 3.353\\
		\hline\hline
	\end{tabular}
	\caption{Wavelengths of the three most intense IR features at 3700~K. *Wavelength shifted to experimental value.}\label{wavelengthFS}
\end{table*}

\end{document}